\documentclass[floatfix,notitlepage,twocolumn]{revtex4-2}%
\usepackage{graphicx} 
\usepackage{amsmath}%
\usepackage{amsfonts}%
\usepackage{amssymb}
\usepackage{bm}
\usepackage{comment}
\usepackage{soul} 
\usepackage{color}
\usepackage[breaklinks]{hyperref}

\def\d{{\partial}}
\def\s{{\sigma}}

\def\k{{ {\bm k} }}

\def\q{{ {\bm q} }}
\def\Q{{ {\bm Q} }}

\def\0{{ {\bm 0} }}

\def\a{{\alpha}}
\def\b{{\beta}}
\def\vp{{\varphi}}
\def\g{{\gamma}}
\def\l{{\lambda}}
\def\r{{ {\bm r} }}
\def\n{{ {\bm n} }}

\def\expo{{ {\rm e} }}
\allowdisplaybreaks[4]

\begin{document}
\title{Topological Spin Hall Effect in Antiferromagnets Driven by Vector N\'eel Chirality} 
\author{ 
Kazuki Nakazawa$^1$, 
Koujiro Hoshi$^1$, 
Jotaro J. Nakane$^2$, 
Jun-ichiro Ohe$^3$, 
Hiroshi Kohno$^2$ 
} 
\address{
$^1$Department of Applied Physics, University of Tokyo, Bunkyo, Tokyo 113-8656, Japan
\\
$^2$Department of Physics, Nagoya University, Nagoya 464-8602, Japan
\\
$^3$Department of Physics, Toho University, 2-2-1 Miyama, Funabashi,  Chiba 274-8510, Japan
}

\date{\today}

\begin{abstract}
Spin Hall effect of spin-texture origin is explored theoretically for antiferromagnetic (AF) metals. It is found that a vector chirality formed by the N\'eel vector gives rise to a topological spin Hall effect. This is topological since it is proportional to the winding number counted by in-plane vector chirality along the sample edge, which can be nonvanishing for AF merons but not for AF skyrmions. The effect is enhanced when the Fermi level lies near the AF gap, and, surprisingly, at weak coupling with small AF gap. These features are confirmed numerically based on the Landauer-B\"uttiker formula. Important roles played by nonadiabatic processes and spin dephasing are pointed out. \end{abstract}

\maketitle

 Spin-charge interconversion has been extensively studied in spintronics with the aim of application to next-generation devices. It is typically achieved by the spin Hall effect (SHE) \cite{SHE0} originating from the relativistic spin-orbit coupling (SOC), mostly in nonmagnetic materials \cite{SHE1,SHE2,SHE3,SHE4,SHE5}. Ferromagnets (FMs) are another class of materials that enable spin-charge conversion, not just as a simple spin source, but also by emergent electromagnetism due to spatiotemporal magnetization dynamics. In particular, a magnetization texture forming a finite scalar spin chirality simulates a magnetic field that affects electrons' orbital motion but in a spin-dependent way. The resulting Hall effect, often called the topological Hall effect (THE), is thus the SHE in essence \cite{com1}. 

Antiferromagnets (AF) are a material having both aspects, magnetic at the microscopic scale but nonmagnetic at the (semi)macroscopic scale, and offers a unique platform to generate pure spin currents. A large SHE was reported in ${\rm Ir_{20}Mn_{80}}$ \cite{Menders2014}, which originates from SOC. Recently, there are some proposals of SHE
that arise from the antiferromagnetic spin texture, providing another means of pure spin-current generation without relying on relativistic SOC. 

In this Letter, we explore theoretically the SHE in AF originating from AF spin textures. From the analogy with FMs, an AF with a textured N\'eel vector ${\bm n}$ is expected to generate a spin Hall current,  
\begin{align}
 \tilde j_{{\rm s},i}^z &= \tilde{\s}_{\rm SH}  \, \n \cdot ( \d_i \n \times \d_j \n )  e E_j , 
\label{eq:js_scalar}
\end{align} 
under an applied electric field $E_j$ ($\tilde{\s}_{\rm SH}$ is a coefficient, and $e>0$ is the elementary charge). 
 Because of the scalar chirality, $\n \cdot ( \d_i \n \times \d_j \n )$, this effect may be termed as a topological spin Hall (TSH) effect \cite{Buhl,Gobel,Akosa}. Such a texture has opposite spin chiralities on different sublattices and deflects spin-up and spin-down electrons in mutually opposite directions; the charge Hall currents then cancel out, but the spin Hall currents add up. However, Eq.~(\ref{eq:js_scalar}) is not a macroscopically observable quantity since its sign depends on the definition of sublattice or ${\bm n}$; it changes sign under ${\bm n} \to -{\bm n}$. We define the physical spin current ${\bm j}_{{\rm s},i}$ through $\tilde{j}_{{\rm s},i}^z = {\bm n} \cdot {\bm j}_{{\rm s},i}$, hence by 
\begin{align}
 j_{{\rm s},i}^\alpha &= \tilde{\s}_{\rm SH}  ( \d_i \n \times \d_j \n )^\alpha  e E_j  . 
\label{eq:js_vector}
\end{align} 
The factor $(\d_i \n \times \d_j \n )^\alpha$ may be identified as an emergent magnetic field in spin channel, and interestingly, it can be expressed as $(\partial_i a_j^\alpha - \partial_j a_i^\alpha )/2$ with an emergent vector potential, 
\begin{align}
 a_i^\alpha = ({\bm n} \times \partial_i {\bm n})^\alpha . 
\label{eq:a_i^alpha}
\end{align} 
This is the vector chirality ($\sim {\bm S}_1 \times {\bm S}_2$ for two spins) formed by the N\'eel vector, and we call it \lq\lq vector N\'eel chirality'' \cite{com:VNC}. Spatially-averaged spin current $\langle j_{{\rm s},i}^\alpha \rangle$ is proportional to a winding number defined by the vector chirality, hence is \lq\lq topological.'' To date, the vector spin chirality is known to induce charge \cite{TT, IN} and (equilibrium) spin currents \cite{TG, KNB, KKAT}, but its AF counterpart in terms of the N\'eel vector has been less focused on.

In the following, we derive Eqs.~(\ref{eq:js_scalar}) and (\ref{eq:js_vector}) and demonstrate the topological character of Eq.~(\ref{eq:js_vector}). The effect is present in systems with AF merons \cite{AFexp3} but not with AF skyrmions \cite{AFSk1,AFSk4,AFSk2,AFSk3,AFexp1,AFexp2,AFSk5}, and is enhanced in the weak-coupling regime. These results are confirmed numerically based on the Landauer-B\"{u}ttiker formula.

\begin{figure}[b]
  \includegraphics[width=85mm]{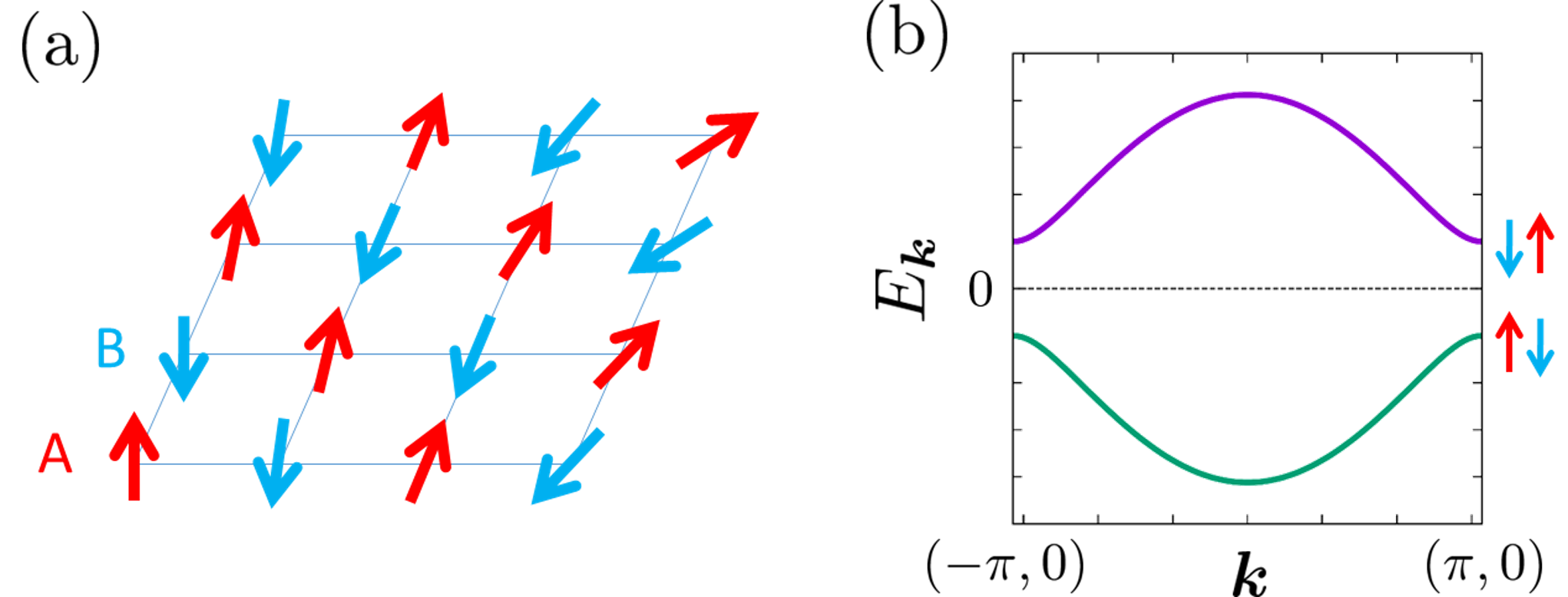}
\vspace*{0mm}
 \caption{
(a) Static magnetic structure considered in this work, a checkerboard type AF on a square lattice with a very slow spatial modulation. The two sublattices (A or B) are indicated by color (red or blue). 
(b) Electron dispersion in a uniform AF state. Each subband is spin degenerate. 
}
\label{fig:AF_structure}
\end{figure}

We consider electrons on a square lattice and coupled to a static spin texture. The Hamiltonian  
\begin{align}
  H &= - t \sum_{(i,j)} c_i^\dagger  c _j  
      - J_{\rm sd} \sum_i {\bm S}_i \cdot ( c_i^\dagger {\bm \sigma} c_i ) 
    + u_{\rm i} {\sum_{i}}' c_i^\dagger c_i  ,
\end{align} 
consists of nearest-neighbor hopping (first term), s-d exchange coupling to localized spins ${\bm S}_i$ (second term), 
and on-site impurity potential (third term), with electron operators $c_i = {}^t ( c_{i\uparrow}, c_{i\downarrow})$ at site $i$ and Pauli matrices ${\bm \sigma} = (\sigma^x, \sigma^y, \sigma^z)$. We assume a slowly-varying checkerboard type AF texture, ${\bm S}_i = S (-1)^i {\bm n}_i$, where ${\bm n}_i$ is the N\'eel vector varying slowly in space [Fig.~\ref{fig:AF_structure} (a)].

With a unitary transformation, $c_i = U_i \tilde c_i$, which diagonalizes the s-d coupling, $U_i^\dagger ({\bm n}_i \cdot {\bm \sigma}) U_i = \sigma^z$, $H$ is transformed into $ H = - t \sum_{(i,j)}  \tilde c_i^\dagger {\rm e}^{iA_{ij}} \tilde c_j - J \sum_{i} (-)^i \tilde c_i^{\dagger} \sigma^z \tilde c_i + u_{\rm i} {\sum_{i}}' \tilde c_i^\dagger \tilde c_i  $, where $J = J_{\rm sd} S$, and $A_{ij}$ is the spin gauge field defined by $U_i^\dagger U_j = \expo^{iA_{ij}}$ \cite{Tatara,Nakane1}. Because of slow variations of the texture, $A_{ij}$ is small and can be treated  perturbatively. The unperturbed state (with $A_{ij}=0$) is a uniform AF, and the electron band splits into spin-degenerate upper and lower bands, $\pm E_{\bm k}$, with an AF gap $2|J|$ in between [Fig.~\ref{fig:AF_structure} (b)]. Here, $E_{\bm k} \equiv \sqrt{\varepsilon_{\bm k}^2 + J^2}$ with $\varepsilon_{\bm k} = -2t(\cos k_x + \cos k_y)$. Also, $A_{ij}$ can be treated in the continuum approximation, $A_{ij} \to A_\mu$, where $\mu$ ($=x,y$) specifies the bond direction of $(i,j)$, and expanded as 
\begin{align}
A_\mu  = \frac{1}{2} A_\mu^\a   \sigma^\a 
 = \frac{1}{2} ( A_\mu^z \sigma^z + {\bm A}_\mu^\perp \cdot {\bm \sigma}^\perp) , 
\label{eq:A}
\end{align}
where $\a = x,y,z$ and $\perp = x,y$. The spin-conserving component $A^z$ describes adiabatic processes, whereas the spin-flip component ${\bm A}^\perp$ induces nonadiabatic transitions. In FM, the latter can be important only in the weak-coupling regime \cite{NBK}, but in AF, both are important because of spin degeneracy of the AF bands. Both produce the same effective field, $(\nabla \times {\bm A}^z )_z = ({\bm A}_x^\perp \times {\bm A}_y^\perp)^z = \n \cdot ( \d_x \n \times \d_y \n )$.

To calculate the spin Hall conductivity, $\sigma_{\rm SH} \equiv \frac{1}{2} (\sigma_{xy}^z - \sigma_{yx}^z)$, we assume a good AF metal and focus on the Fermi-surface contribution \cite{Streda}, 
\begin{align}
 \sigma_{ij}^z (\Q) 
&= - \frac{e\hbar}{4\pi} {\rm Tr} 
    \left\langle {\cal J}_{{\rm s}, i}^z G_{\k_{+},\k'}^{\rm R} 
                     {\cal J}_j  G_{\k',\k_{-}}^{\rm A} \right\rangle_{\rm i} , 
\label{eq:conRA}
\end{align} 
where ${\cal J}_{{\rm s}, i}^z$ and ${\cal J}_j$ are spin-current and number-current vertices, $\rm Tr$ means the trace in spin, sublattice, and $\k$ spaces ($\k$, $\k'$-integrals), and $\langle \cdots \rangle_{\rm i}$ represents impurity average. The Green's function $G_{\k,\k'}^{\rm R(A)} = (\mu - H \pm i0)_{\k, \k'}^{-1} $ takes full account of impurities and the gauge field, and $\k_\pm = \k \pm \Q/2$. We treat the impurity scattering in the Born approximation with ladder vertex corrections (VC) \cite{suppl}. The superscript $z$ on $\sigma_{ij}^z$ and ${\cal J}_{{\rm s}, i}^z$ indicates the spin component in the rotated frame, thus it is the component {\it projected to the local N\'eel vector} ${\bm n}$.

After a standard procedure (see Supplemental Material \cite{suppl}), we obtain Eq.~(\ref{eq:js_scalar}) with $\tilde \sigma_{\rm SH} = \tilde \sigma_{\rm SH}^{(0)} + \tilde \sigma_{\rm SH}^{(1)}$, 
\begin{align}
 \tilde \sigma_{\rm SH}^{(0)} 
&=  (J \tau)^2 \frac{t^2 \nu}{\mu}  \left( 1 - \frac{J^2}{\mu^2} \right) C_{xy} , 
\label{eq:thc-wov}
\\
 \tilde \sigma_{\rm SH}^{(1)} 
&= (J \tau)^2 \frac{t^2 \nu}{\mu} \frac{8t^2}{\mu^2 + J^2} 
      \left( \frac{\tau^{-1}}{\tau_\varphi^{-1} + \tau_{\rm s}^{-1} } \right)   C_{xx}^2  , 
\label{eq:thc-wv}
\end{align} 
where $\tilde \sigma_{\rm SH}^{(0)}$ is the contribution without VC, which comes from both adiabatic and nonadiabatic processes, and $\tilde \sigma_{\rm SH}^{(1)}$ is the contribution with VC, coming only from  nonadiabatic processes. Here, $\nu = \nu(\mu)$ is the density of states (per spin) at chemical potential $\mu$, $C_{ij} = \langle 1 - \cos k_i \cos k_j \rangle_{\rm FS}$ is the Fermi surface average \cite{com:FS}, $\tau = [ \g_0 + (J/\mu) \g_3 ]^{-1} /2$ is the scattering time ($\g_0 = \pi n_{\rm i} u_{\rm i}^2 \nu$ and $\g_3 = (J/\mu)\g_0$ are the sublattice-independent and dependent parts, respectively, of the damping, and $n_{\rm i}$ is the impurity concentration), and 
\begin{align}
 \frac{1}{\tau_\varphi} 
 = \frac{4J}{\mu} \frac{\mu^2 + J^2}{\mu^2 - J^2} \g_3 
 = \frac{2J^2}{\mu^2 - J^2} \frac{1}{\tau}  , 
\label{eq:tau_phi}
\end{align} 
is the \lq\lq spin dephasing'' rate \cite{Manchon}. We introduced a finite spin relaxation rate $\tau_{\rm s}^{-1}$ by hand \cite{com0}; without $\tau_{\rm s}^{-1}$, we would have an unphysical result that $\tilde \sigma_{\rm SH}^{(1)}$ does not vanish in the limit $J \to 0$. Note that $\tau_\varphi^{-1}$ differs from $\tau_{\rm s}^{-1}$ in that it does not require spin-dependent scattering, randomizes only the transverse ($\perp {\bm n}$) components of the electron spin (see below), and vanishes as $J \to 0$. The results (\ref{eq:thc-wov}) and (\ref{eq:thc-wv}) are obtained at the leading order, i.e., second order in spatial gradient and second order in $\tau$.

\begin{figure}[t]
\hspace*{-6mm}
  \includegraphics[width=85mm]{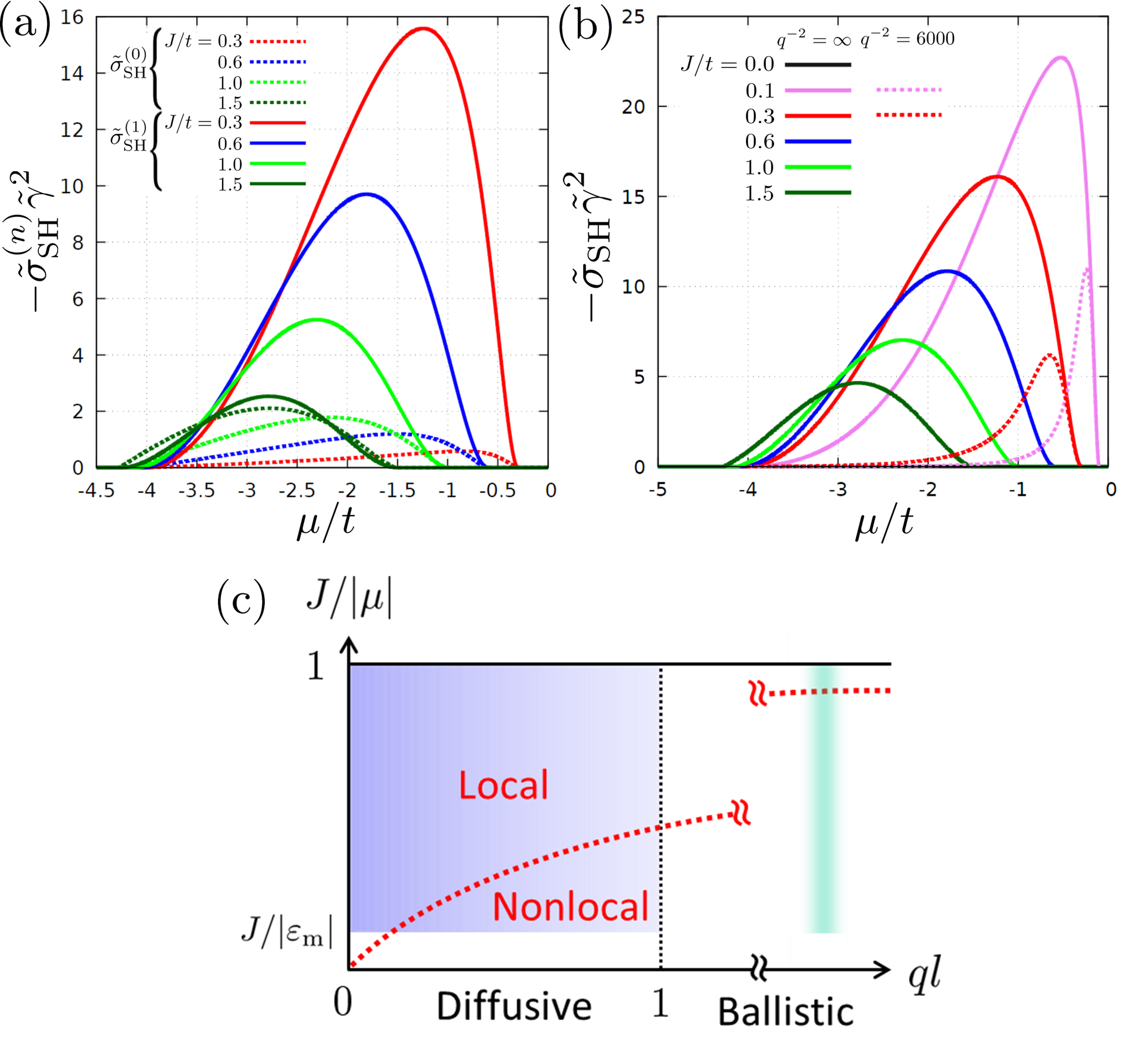}
\vspace*{0mm}
 \caption{(a,b)  Normalized topological spin Hall conductivity vs. chemical potential $\mu$ for several choices of $J/t$. 
(a) $\tilde \sigma_{\rm SH}^{(0)} \tilde \gamma^2$ and $\tilde \sigma_{\rm SH}^{(1)} \tilde \gamma^2$, where $\tilde \gamma = \pi n_{\rm i} u_{\rm i}^2 /t^2$ is a dimensionless damping parameter. 
(b) $\tilde \sigma_{\rm SH} = \tilde \sigma_{\rm SH}^{(0)} + \tilde \sigma_{\rm SH}^{(1)}$. In (b), $\tilde \sigma_{\rm SH}$ with finite $q$ are also shown (dotted lines). These are odd functions of $\mu$, hence plotted only for the lower AF band. The parameters used are $\tilde \gamma = 0.2$ and $\tau_{\rm s}^{-1} = 10^{-4}t$. 
(c) Characteristic parameter regions for the TSH conductivity. The red dashed line, given by $J/|\mu| = ql / \sqrt{4 + (ql)^2}$ in the diffusive regime, is a crossover line separating the local and nonlocal field regions, and $\varepsilon_{\rm m} = \sqrt{(4t)^2 + J^2}$. The analytical results, Eqs.~(\ref{eq:thc-wov}), (\ref{eq:thc-wv}), and (\ref{eq:diff}), apply to the blue shaded region, while the numerical results (Fig.~\ref{fig:LB}) apply to the green shaded region. 
}
\label{fig:TSHE_plot}
\end{figure}

\begin{figure}[t]

  \includegraphics[width=88mm]{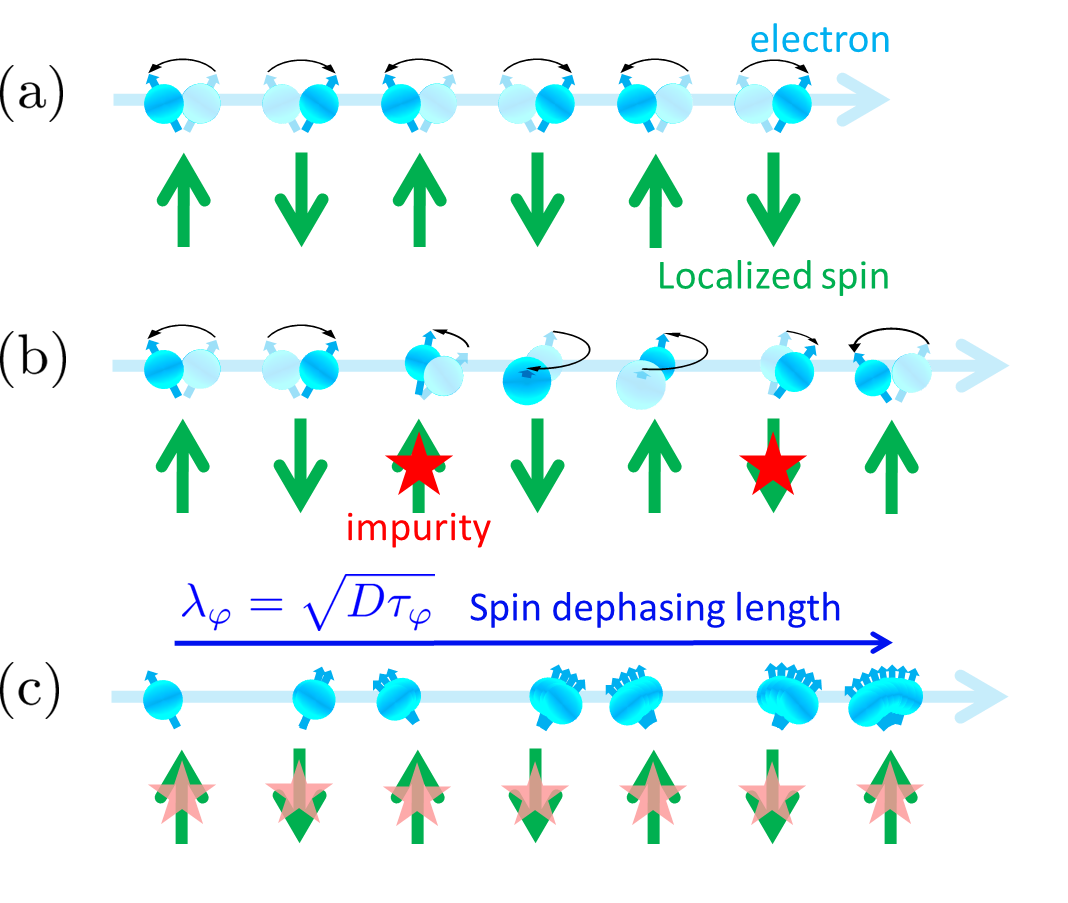}
\vspace*{-9mm}
 \caption{
Physical picture of electron spin transport in a uniform antiferromagnet. The blue sphere with an arrow represents an electron, the green arrow a localized spin, and the red star a nonmagnetic impurity. 
 (a) The electron spin precesses around the local moment, alternating its sense from site to site. 
 (b) Interaction with impurities locally modifies the precession. 
 (c) A \lq\lq collective'' transverse spin density contributed from many electrons decays and loses its original information through the impurity scattering. This is because the degree of the modification, mentioned in (b), varies 
from electron to electron. This is called \lq\lq dephasing'' and the characteristic length is the \lq\lq dephasing length'' $\lambda_\varphi = \sqrt{D \tau_\varphi}$. The orange stars represent averaged impurities.
}  
\label{fig:picture}
\end{figure}

The coefficients $\tilde \sigma_{\rm SH}^{(0)}$ and $\tilde \sigma_{\rm SH}^{(1)}$ are plotted  in Fig.~\ref{fig:TSHE_plot} (a). They are comparable in magnitude at large $J$ ($\sim 1.5 t$), but as $J$ is reduced, $\tilde \sigma_{\rm SH}^{(1)}$ grows markedly whereas $\tilde \sigma_{\rm SH}^{(0)}$ decreases. The sum $\tilde \sigma_{\rm SH} = \tilde \sigma_{\rm SH}^{(0)} + \tilde \sigma_{\rm SH}^{(1)}$ is plotted in Fig.~\ref{fig:TSHE_plot} (b) by solid lines, which grows as $J$ is reduced, especially near the AF gap edge, but finally vanishes at $J=0$. Since $\tilde \sigma_{\rm SH}^{(1)}$ comes solely from nonadiabatic processes, these results show that the combined effect of nonadiabaticity and the VC is important for the present SHE \cite{com2}. Physically, a nonadiabatic process produces a transverse spin polarization, and the VC describes its collective transport, which is however limited by spin dephasing \cite{Manchon,Nakane1,HLN}. The origin of the enhancement at small $J$ can be traced to the reduced dephasing at small $J$ \cite{suppl}. As seen from Eq.~(\ref{eq:tau_phi}), the spin dephasing arises through $\g_3$, a sublattice asymmetry in (nonmagnetic) scattering \cite{Nakane1,Nakane2}, and its physical picture is illustrated in Fig.~\ref{fig:picture}.

The obtained result, Eq.~(\ref{eq:js_scalar}), needs to be interpreted with care. It arises with the scalar chirality formed by the N\'eel vector ${\bm n}$, and changes sign under ${\bm n} \to - {\bm n}$. This is not a pleasant situation since any physical quantity measurable by (semi)macroscopic means should not depend on the sign of ${\bm n}$, or on the definition of sublattice. This (apparent) puzzle is resolved if we note that the spin component of the calculated spin current $\tilde{j}_{x,{\rm s}}^z$ is the one projected to the N\'eel vector $\bm n$. Therefore, we write $\tilde{j}_{\rm s}^z = {\bm n} \cdot {\bm j}_{\rm s}$ and identify ${\bm j}_{\rm s}$ as a physical spin current. The physical spin Hall current is thus given by Eq.~(\ref{eq:js_vector}).

It is in fact possible to obtain Eq.~(\ref{eq:js_vector}) directly. By assuming $J$ is small and treating it perturbatively, we found the spin current arises at second order in $J$ \cite{suppl}, 
\begin{align} 
 j_{{\rm s},i}^\alpha 
&=  (J \tau)^2  \frac{t^2 \nu}{\mu}  C_{xy} 
      (\partial_i {\bm n} \times \partial_j {\bm n} )^\alpha e E_j . 
\label{eq:js_pert}
\end{align}
This contrasts with the THE in FM caused by scalar spin chirality, which starts at third order ($\sim J^3$) \cite{Tatara2002}, and demonstrates that the essential quantity for the present SHE is the vector (not scalar) chirality. That Eq.~(\ref{eq:js_pert}) is an even function of $J$ (or $J {\bm n}$) is consistent with the fact that the spin current is even under time reversal.

The expression Eq.~(\ref{eq:js_vector}) holds locally in space (as far as the variation of ${\bm n}$ is sufficiently slow). As a spin current measured experimentally, we consider a spatially-averaged one, $\langle {\bm j}_{\rm s}^\alpha \rangle = \Omega^{-1} \int {\bm j}_{\rm s}^\alpha dx dy$ (in two dimensions), where $\Omega$ is the sample area. It can be written as  
\begin{align}
 \langle {\bm j}_{\rm s}^\alpha \rangle 
&= \pi  \tilde{\s}_{\rm SH} \frac{m^\alpha}{\Omega} \left( e {\bm E} \times \hat{z} \right) , 
\label{eq:js_integrated}
\end{align} 
where
\begin{align}
 m^\alpha &= \frac{1}{2\pi} \int (\nabla \times {\bm a}^\alpha )_z  dxdy 
= \frac{1}{2\pi} \oint {\bm a}^\alpha \cdot d {\bm \ell}  , 
\end{align} 
and $ a_i^\alpha = ({\bm n} \times \partial_i {\bm n})^\alpha$ [Eq.~(\ref{eq:a_i^alpha})] is an emergent vector potential in spin channel. The line integral is taken along the sample perimeter. If the system has easy-plane magnetic anisotropy, and the N\'eel vector on the sample edge is constrained to lie in-plane, e.g., $x$-$y$ plane, the line integral of the vector chirality defines a topological winding number $m^z \in \mathbb{Z}$ in $\pi_1(S^1)$. The spin Hall conductivity is thus proportional to the topological number density $m^z/\Omega$, and this fact resurrects the naming \lq\lq topological'' spin Hall effect. We emphasize that it is characterized by the vector chirality of N\'eel vector along the sample edge. Therefore, the present TSHE is absent for AF skyrmions, in which the N\'eel vector at the edge is uniaxial. On the other hand, it is finite for AF merons, which have finite in-plane winding of the N\'eel vector along the edge.

\begin{figure}[t]
  \includegraphics[width=85mm]{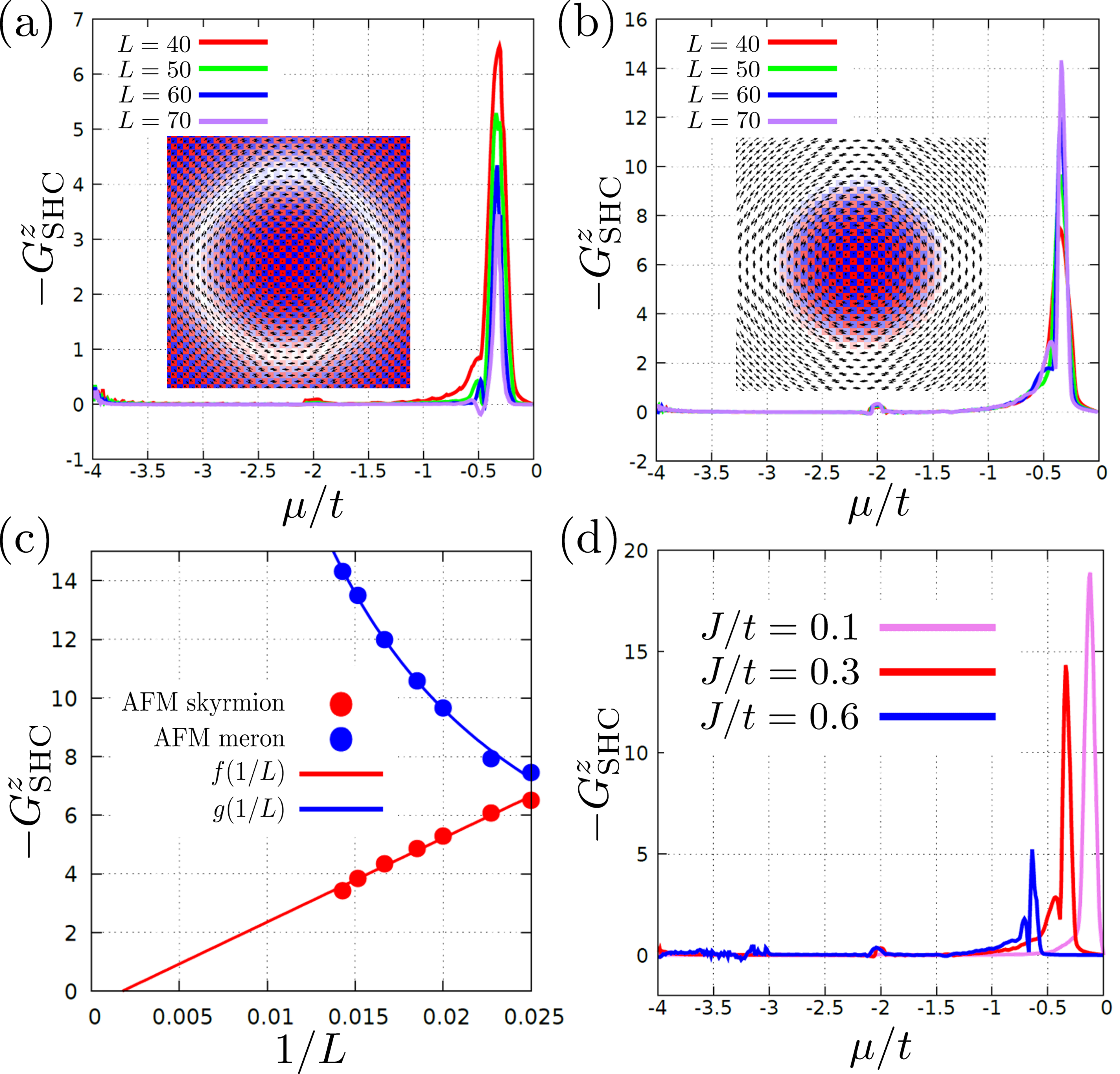}
 \caption{
Topological spin Hall conductance $(G_{\rm SHC}^z)$ based on the  
Landauer-B\"uttiker formula for finite systems with $L \times L$ sites.  
(a) AF skyrmion system. 
(b) AF meron system. 
(c) $L$-dependence of the peak value of $G_{\rm SHC}^z$. The data are fitted with functions, $f(x) = 286x - 0.504$ and $g(x)=0.237/x - 2.22$. 
(d) AF meron system with $L=70$ for several choices of $J/t$. We took $J/t = 0.3$ [except in (d)] and meron/skyrmion radius  $r=15$. The data are symmetrized with respect to $J \to -J$, as explained in \cite{suppl}. 
}
\label{fig:LB}
\end{figure}

To verify these results, we have conducted numerical works based on the four-terminal Landauer-B\"uttiker formula \cite{Ohe2017}. We consider ballistic systems with $L \times L$ sites without disorder, and containing a single AF skyrmion or a single AF meron. For both textures, the spin Hall conductance $G_{\rm SHC}^z$ shows a strong peak just below the AF gap [Fig.~\ref{fig:LB} (a) and (b)], which, however, behave oppositely as $L$ is increased (with the skyrmion/meron size fixed); for the AF skyrmion the peak decreases with $L$ and seems to vanish in the thermodynamic limit. In contrast, for the AF meron it increases with $L$ [Fig.~\ref{fig:LB} (c)]. This is consistent with the analytical result, which is valid for infinite-size systems. Plots for several $J/t$ values are shown in Fig.~\ref{fig:LB} (d)  for the AF meron system, showing that it is indeed enhanced at small $J/t$. All these agree with the analytic results, except for the detailed shape of $\mu$-dependence. 

The discrepancy in shape ($\mu$-dependence) between the numerical [Fig.~\ref{fig:LB} (d)] and analytic results [Fig.~\ref{fig:TSHE_plot} (b)] may be understood as due to the nonlocality effect in the former. To illustrate this, let us first consider the diffusive regime. As the typical wave number $q$ of the N\'eel texture (i.e., inverse of meron/skyrmion size) is increased, Eq.~(\ref{eq:thc-wv}) is modified as 
\begin{align}
 (\tau_\varphi^{-1} + \tau_{\rm s}^{-1} )^{-1} \to 
 (\tau_\varphi^{-1} + \tau_{\rm s}^{-1} +  D q^2  )^{-1} , 
\label{eq:diff}
\end{align} 
in the denominator, where $D$ is the diffusion constant. When electron spin diffusion ($Dq^2$) occurs faster than spin dephasing ($\tau_\varphi^{-1}$), the effective field becomes \lq\lq nonlocal''. Similar feature has been noted for FMs, in which $Dq^2$ is compared to the exchange splitting \cite{NBK}. Here in AF, it is compared to the (much smaller) spin dephasing rate, $\tau_\varphi^{-1} $, hence the present SHE enters the nonlocal regime rather easily compared to the THE in FM. More explicitly, the nonlocality appears if 
\begin{align} 
 q l > \frac{2J}{\sqrt{\mu^2 - J^2}} , \ \ \  {\rm or}  \ \ \  
 | \mu | > J \sqrt{1 + ( 2/ql )^2} , 
\label{eq:local}
\end{align} 
where $l$ is the mean free path. In Fig.~\ref{fig:TSHE_plot} (b), the analytic results with $q^{-2} = 6000$ (with lattice constant taken unity) are plotted by dotted lines. The suppression due to nonlocality is more significant at larger $|\mu|$ (away from the AF gap), leaving a sharp peak in the vicinity of the AF gap edge. Since cleaner systems enter the nonlocal regime more easily [see Eq.~(\ref{eq:local}) and a red dotted line in Fig.~\ref{fig:TSHE_plot} (c)], this feature is expected to persist into the ballistic regime with a wider nonlocality region. The shape of Fig.~\ref{fig:LB} (d) may thus be understood as due to the nonlocality effect.

Thus, as in the case of THE in FM \cite{NBK}, the present TSHE in AF exhibits various characteristic regimes [Fig.~\ref{fig:TSHE_plot} (c)]. These are summarized as follows. First, for a ballistic and local regime, the effect is truly topological. As $q$ is increased and the nonlocal effects become important, the SHC deviates from the topological expression. In the diffusive case, it is difficult to have the topological expression because of dephasing (and nonlocality), but the effect is enhanced for weak-coupling AF with small AF gap. An interesting possibility may be found in mesoscopic systems, for which the effect can be topological even if the system is in a diffusive regime. 

The emergent vector potential ${\bm a}^\alpha$ in the spin channel, identified here through TSHE, has more generality. In a study on THE in canted AF \cite{Nakane1}, an emergent vector potential in charge channel was identified as $l^\alpha {\bm a}^\alpha$, where $l^\alpha$ is the canting (uniform) moment. Also, ${\bm a}^\alpha$ can be expressed as $a_i^\alpha = - ({\cal R} {\bm A}^\perp_i)^\alpha$ \cite{Kohno2007}, where ${\cal R}$ is an SO(3) matrix that connects the rotated and the original frames (e.g., ${\bm n} = {\cal R} \hat z$), showing its conformity with the spin gauge field ${\bm A}^\perp_i$. These facts reinforce our interpretation of ${\bm a}^\alpha$ as an effective vector potential in spin channel.

To realize the present TSHE experimentally, a prime candidate texture is ${\bm n}$-meron. Such texture was found very recently in $\alpha$-Fe$_2$O$_3$ \cite{AFexp3}, which is however an insulator; search for metallic systems is desired. Another candidate is a canted AF; if the ferromagnetic moment ${\bm l}$ (due to canting) forms a skyrmion (called \lq\lq ${\bm l}$-skyrmion'' in \cite{Nakane1}), topological consideration shows that the N\'eel vector winds at least twice around the skyrmion, i.e., $m^z = 2$ per skyrmion  \cite{Nakane1,Meshcheriakova}. A recent experiment on thin films of Ce-doped CaMnO$_3$, a canted AF, observed skyrmion bubbles formed by the (weak) {\it ferromagnetic} moment \cite{Vistoli}. Therefore, this system can also be a candidate for the present TSHE.

Finally, we discuss the relationship with previous theoretical studies. In Ref.~\cite{Gobel}, the TSH conductivity was investigated in an AF skyrmion lattice with a focus on the intrinsic (Berry curvature) contribution. It is a future issue to investigate such intrinsic contribution in our framework. For example, one may consider a \lq\lq meron-antimeron lattice'' which contains a vortex with $m = \pm 2$ per unit cell. In Ref.~\cite{Akosa}, the Landauer-B\"uttiker method was used to study the TSHE in AF skyrmion systems, and a finite TSHE was found for {\it finite-size systems}, which does not contradict with our result because of finite size. More importantly, an increase of the spin Hall conductivity was pointed out for special impurity configurations, and it is also interesting to investigate how the impurity configuration affect the spin dephasing and TSHE.

To summarize, we have studied a spin Hall effect due to magnetic textures in AF metals. By analytic calculations, we found a topological contribution proportional to the winding number defined by vector chirality. This is finite for AF merons but not for AF skyrmions, and is enhanced at weak coupling. These results may provide hints to enhance spin currents and give directions for experimental investigations and device fabrications. The results are confirmed by numerical calculations based on the Landauer-B\"{u}ttiker formula. Important roles played by nonadiabatic processes and spin dephasing are pointed out.

We would like to thank A. Yamakage, J. Fujimoto, T. Yamaguchi, Y. Imai,  A. Matsui, and T. Nomoto for valuable discussions. 
This work was partly supported by JSPS KAKENHI Grant Numbers JP15H05702, JP17H02929,  
JP19K03744 and No. JP21H01799, and the Center of Spintronics Research Network of Japan. 
KN is supported by JSTCREST (JP-MJCR18T2) and JSPS KAKENHI Grant Number JP21K13875.  
JJN  is  supported  by  a  Program  for  Leading  Graduate Schools 
\lq\lq Integrative Graduate Education and Researchin  Green  Natural  Sciences'' 
and  Grant-in-Aid  for  JSPS Research Fellow Grant Number JP19J23587.

\newpage

\newpage

\onecolumngrid
\clearpage
\begin{center}
\textbf{\large Supplemental material \\ Topological Spin Hall Effect in Antiferromagnets Driven by Vector N\'eel Chirality}
\end{center}

\setcounter{equation}{0}
\setcounter{figure}{0}
\setcounter{table}{0}
\setcounter{page}{1}
\makeatletter
\renewcommand{\theequation}{S\arabic{equation}}
\renewcommand{\thefigure}{S\arabic{figure}}
\renewcommand{\thetable}{S\arabic{table}}

\section{Preliminary}
\label{sec:basics}

\subsection{Green's function and vertex corrections} 
\label{sec:GrF}
\begin{figure}[b]
 \includegraphics[width=140mm]{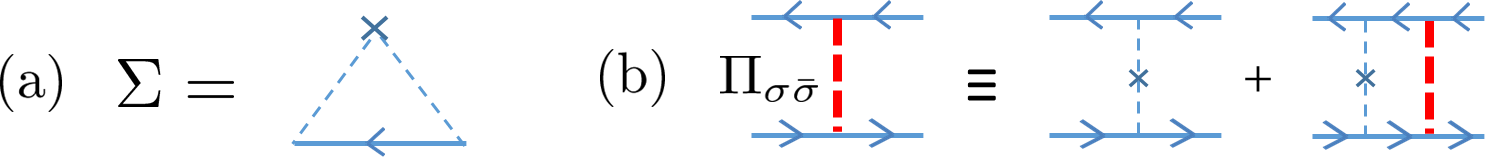}
\vspace{-0mm}
 \caption{
Feynman diagrams for the self energy (a) and the ladder-type vertex correction (b). The blue solid line with arrow is a retarded or advanced Green's function. The blue broken line with a cross represents (averaged) impurity scattering.
}
 \label{fig:se_vc}
\end{figure}

 Before presenting the calculation of the topological spin Hall conductivity (TSHC), we define some building blocks of the calculation. The first one is the Green's function that does not contain the spin gauge field but with the effects of impurities evaluated in the Born approximation [Fig.~\ref{fig:se_vc}(a)].   
\begin{align}
G_\k^{\rm R} (\varepsilon) = \mu^{\rm R} (1 \otimes 1) 
   + J^{\rm R} (\sigma^z \otimes \tau_3) + T_\k^{\rm R} (1 \otimes \tau_1) , 
\label{eq:GrF}
\end{align}
where $\mu^{\rm R} = (\varepsilon + \mu + i\gamma_0)/D_{\bm k}^{\rm R}$, $J^{\rm R} = (i\gamma_3 - J)/D_{\bm k}^{\rm R}$, $T_{\bm k}^{\rm R} = T_\k / D_{\bm k}^{\rm R} = -2t (\cos k_x + \cos k_y)/D_{\bm k}^{\rm R}$, with $D_{\bm k}^{\rm R} (\varepsilon) = (\varepsilon + \mu)^2 - E_{\bm k}^2 + 2i \left[ (\varepsilon + \mu)\gamma_0 + J\gamma_3 \right]$, and $E_{\bm k} = \sqrt{T_{\bm k}^2 + J^2}$. The damping constants are given by $\g_0 = \pi n_{\rm i} u_{\rm i}^2 \nu$ and $\g_3 = (J/\mu)\g_0$, where $n_{\rm i}$ is the impurity concentration and $\nu = (1/N) \sum_\k \delta(|\mu| - E_\k)$ is the density of states per spin at the chemical potential $\mu$, and  we define elastic scattering (mean free) time $\tau$ by $(2\tau)^{-1} = \g_0 + (J/\mu)\g_3 =  (1 + J^2/\mu^2) \g_0$. In Eq.~(\ref{eq:GrF}), the spin degree of freedom is described by the Pauli matrices ${\bm \s} = (\s^x, \s^y, \s^z)$, and the sublattice degree of freedom by another Pauli matrices ${\bm \tau} = (\tau_1, \tau_2, \tau_3)$. We occasionally suppress/simplify the wave-vector dependence, such as $G = G_\k$, $G_\pm = G_{\k \pm \q/2}$, $G^{\rm R(A)} \equiv G_\k^{\rm R(A)}$, or $G_\pm^{\rm R(A)} \equiv G_{\k \pm \q/2}^{\rm R(A)}$.

To be consistent with the self-energy considered above, we also consider the ladder-type vertex corrections [Fig.~\ref{fig:se_vc}(b)]. The particle-hole correlation with opposite spins (i.e., transverse spin diffusion propagator) plays an essential role in the present TSHE. Explicit form of the spin diffusion propagator (red broken line in Fig. \ref{fig:wvc}) is  \cite{Nakane, Manchon} 
\begin{align}
 \Pi^{(\a 0 )(\a 0)} (\q ,\omega)
&= \frac{4}{\pi \nu \tau^2} \frac{\mu^2}{\mu^2 - J^2} 
   \frac{1}{ \tau_{\rm \varphi}^{-1} + \tau_{\rm s}^{-1} + Dq^2 - i\omega } 
  \equiv \Pi_{\bar\sigma \sigma} ,  
\label{eq:VC}
\\
 \Pi^{(\a 3 )(\a 3)} (\q ,\omega) &= \frac{4}{\pi \nu \tau} \frac{\mu^2}{\mu^2 + J^2}  , 
\end{align}
where $\a = x,y$, $D = \frac{\mu^2 - J^2}{\mu^2}\langle (2t \sin k_x)^2 \rangle_{\rm FS}$ is the diffusion constant, $\tau_\varphi$ is the spin dephasing time [Eq.~(9) in the main text], $\tau_{\rm s}$ is the (isotropic) spin relaxation time, and $\Pi^{(\a 0)(\a 0)}$ corresponds to Eq.~(13) in the main text. These have been derived in Ref.~\cite{Nakane}. 

\subsection{Current and density vertices}
\label{sec:VV}

 The vertices we use in the following are defined as follows,  
\begin{align}
 j_i &\equiv 2t \sin k_i \sigma^0 \tau_1,
\\
 j_i^\alpha &\equiv 2t \sin k_i \sigma^\alpha \tau_1,
\\
\rho_{ij}^{\alpha} &\equiv 2t \cos k_i \sigma^\alpha \tau_1 \delta_{ij}, 
\\
\rho_{ij}^{\a \b} &\equiv 2t \cos k_i \sigma^\a  \sigma^\b \tau_1 \delta_{ij} + {\rm h.c.}, 
\end{align}
where $\alpha,\beta=x,y,z$. 
 We write as $j_{i \pm} \equiv 2et \sin ( k_i \pm q_i/2 ) \sigma^0 \tau_1$ and 
$j_{i \pm}^\a \equiv 2t \sin ( k_i \pm q_i/2 ) \sigma^\a \tau_1$. 
 The spin-current and number-current vertices are given by 
\begin{align}
 {\cal J}_{{\rm s}, i}^\alpha &= j_i^\alpha + \rho_{ij}^{\a \b} A_j^\beta  ,  
\\
 {\cal J}_i &= j_i + \rho_{ij}^{\alpha} A_j^\alpha  . 
\end{align}

\section{Calculation of TSHC without VC}
\label{sec:w/o-VC}

\begin{figure}
 \includegraphics[width=150mm]{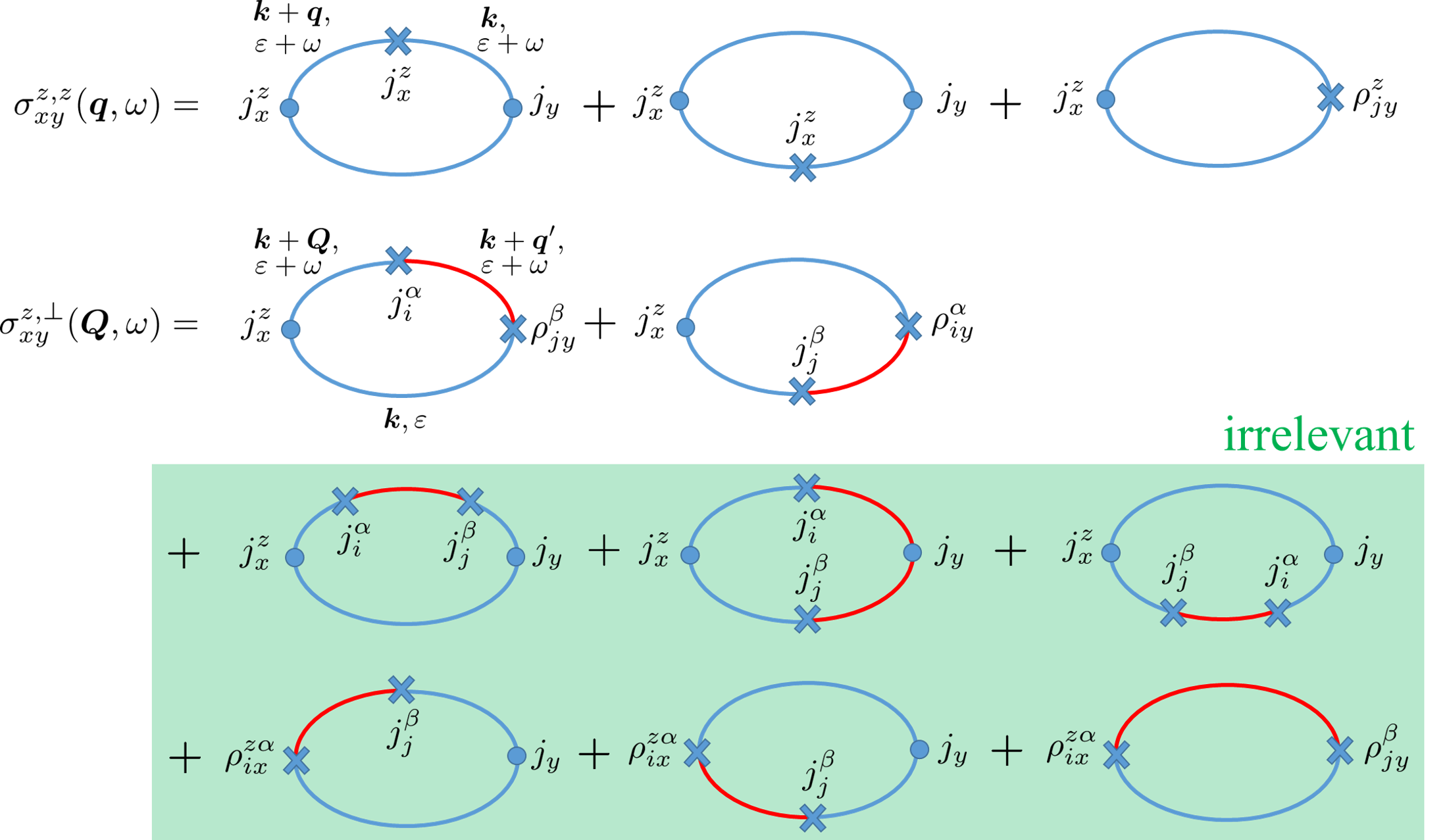}
\vspace{-0mm}
 \caption{
Feynman diagrams for the TSH conductivity without vertex corrections. The blue and red solid lines represent the Green's functions with mutually opposite spin states. The total wave vector ${\bm Q} = \q + \q'$ in the nonadiabatic terms ($\sigma_{xy}^{z,\perp}$) is provided by the gauge fields, which will be set $\Q \to {\bm 0}$ later in the calculation. The diagrams in the green-shaded region turned out to vanish because of symmetry. 
}
 \label{fig:wovc}
\end{figure}

 We first study the contribution without vertex corrections (VC). After a perturbative treatment of the gauge field and taking impurity averaging, we divide the perturbation terms into two parts; adiabatic part (which contains $A^z$ only) and nonadiabatic part (which contains $A^\perp$ only). The adiabatic contribution consists of three terms, 
\begin{align}
\sigma_{xy}^{z,z} = -\frac{e}{4\pi} \cdot \frac{1}{2} A_i^z  
\sum_{\bm k} {\rm tr} \left[ 
 j_x^z G_+^{\rm R} j_i^z G_-^{\rm R} j_{y-} G_-^{\rm A} 
+ j_x^z G_+^{\rm R} j_{y-} G_+^{\rm A} j_i^z G_-^{\rm A} 
+ j_x^z G_+^{\rm R} \rho_{iy}^z G_-^{\rm A} \right] ,
\label{eq:thc-ad}
\end{align} 
and the nonadiabatic contribution consists of eight terms,  
\begin{align}
\sigma_{xy}^{z,\perp} = \sigma_{xy}^{z,1a} + \sigma_{xy}^{z,1b} + \sigma_{xy}^{z,1c} + \sigma_{xy}^{z,2a} + \sigma_{xy}^{z,2b} + \sigma_{xy}^{z,3a} + \sigma_{xy}^{z,3b} + \sigma_{xy}^{z,4} , 
\end{align} 
where
\begin{align}
\sigma_{xy}^{z,1a} &= -\frac{e}{4\pi} \cdot \frac{1}{4} A_i^\alpha A_j^\beta \sum_{\bm k} {\rm tr} \left[ j_x^z G^{\rm R} j_i^\alpha G^{\rm R} j_j^\beta G^{\rm R} j_y G^{\rm A} \right]  , 
\\
\sigma_{xy}^{z,1b} &= -\frac{e}{4\pi} \cdot \frac{1}{4} A_i^\alpha A_j^\beta \sum_{\bm k} {\rm tr} \left[ j_x^z G^{\rm R} j_i^\alpha G^{\rm R} j_y G^{\rm A} j_j^\beta G^{\rm A} \right]  , 
\\
\sigma_{xy}^{z,1c} &= -\frac{e}{4\pi} \cdot \frac{1}{4} A_i^\alpha A_j^\beta \sum_{\bm k} {\rm tr} \left[ j_x^z G^{\rm R} j_y G^{\rm A} j_i^\alpha G^{\rm A} j_j^\beta G^{\rm A} \right]  , 
\\
\sigma_{xy}^{z,2a} &= -\frac{e}{4\pi} \cdot \frac{1}{4} A_i^\alpha A_j^\beta \sum_{\bm k} {\rm tr} \left[ \rho_{ix}^{z \alpha} G^{\rm R} j_j^\beta G^{\rm R} j_y G^{\rm A} \right]  , 
\\
\sigma_{xy}^{z,2b} &= -\frac{e}{4\pi} \cdot \frac{1}{4} A_i^\alpha A_j^\beta \sum_{\bm k} {\rm tr} \left[ \rho_{ix}^{z \alpha} G^{\rm R} j_y G^{\rm A} j_j^\beta G^{\rm A} \right]  , 
\\
\sigma_{xy}^{z,3a} &= -\frac{e}{4\pi} \cdot \frac{1}{4} A_i^\alpha A_j^\beta \sum_{\bm k} {\rm tr} \left[ j_x^z G^{\rm R} j_i^\alpha G^{\rm R} \rho_{jy}^\beta G^{\rm A} \right]  , 
\\
\sigma_{xy}^{z,3b} &= -\frac{e}{4\pi} \cdot \frac{1}{4} A_i^\alpha A_j^\beta \sum_{\bm k} {\rm tr} \left[ j_x^z G^{\rm R} \rho_{jy}^\beta G^{\rm A} j_i^\alpha G^{\rm A} \right]  , 
\\
\sigma_{xy}^{z,4} &= -\frac{e}{4\pi} \cdot \frac{1}{4} A_i^\alpha A_j^\beta \sum_{\bm k} {\rm tr} \left[ \rho_{ix}^{z \alpha} G^{\rm R} \rho_{jy}^\beta G^{\rm A} \right]  . 
\end{align}
 The corresponding Feynman diagrams are shown in Fig. \ref{fig:wovc}. 

 Most of the nonadiabatic terms can be disregarded, however. First, $\sigma_{xy}^{z,1a}, \sigma_{xy}^{z,1b}$, and $\sigma_{xy}^{z,1c}$ do not contain anti-symmetric components, so they are not studied here. Also, we find $\sigma_{xy}^{z,2a} = \sigma_{xy}^{z,2b} = 0$ and $\sigma_{xy}^{z,4} = 0$ because of the cancellation with the Hermitian conjugate part of $\rho_{ij}^{z \a}$. Hence, it is sufficient to consider $\sigma_{xy}^{z,\perp} = \sigma_{xy}^{z,3a} + \sigma_{xy}^{z,3b}$, or 
\begin{align}
\sigma_{xy}^{z,\perp} = -\frac{e}{4\pi} \cdot \frac{1}{4} A_i^\alpha A_j^\beta \sum_{\bm k} 
  {\rm tr} \left[ j_x^z G^{\rm R} j_i^\alpha G^{\rm R} \rho_{jy}^\beta G^{\rm A} 
                   + j_x^z G^{\rm R} \rho_{jy}^\beta G^{\rm A} j_i^\alpha G^{\rm A} \right] .
\label{eq:thc-nad}
\end{align} 

 Let us first calculate the adiabatic contribution, Eq.~(\ref{eq:thc-ad}), which is expressed as. 
\begin{align}
\sigma_{xy}^{z,z} 
&= -\frac{e}{4\pi} A_i^z  
\sum_{\bm k} {\rm tr} 
\left[ G_{{\bm k} -}^{\rm A} j_x^z G_{{\bm k} + }^{\rm R} 
\left\{  j_i^z \left( G^{\rm R} j_y - G^{\rm A} j_y \right)_-  - \left(  j_y G^{\rm R} - j_y G^{\rm A} \right)_+ j_i^z \right\} \right] . 
\label{eq:thc-ad2}
\end{align} 
 Expanding it with respect to $q$, we write
\begin{align}
\sigma_{xy}^{z,z} = -\frac{e}{16 \pi} q_j A_i^z \left( K_{ij} - K_{ji} \right) 
			  = -\frac{e}{16 \pi} \left( \q \times {\bm A}^z \right)_z K  , 
\label{eq:adiabatic}
\end{align}
where
\begin{align}
K_{ij} &= \sum_{\bm k} {\rm tr} 
\left[ 
G^{\rm A} j_x^z G^{\rm R} j_i^z \left( G^{\rm R} - G^{\rm A} \right) \rho_{jy}^0
\right]
= \delta_{ix} \delta_{jy} K ,
\\
K &= (2t)^3 \sum_{\bm k} \sin^2 k_x \cos k_y \cdot  {\rm tr} \left[ \sigma^z \tau_1 G^{\rm R} \sigma^z \tau_1 G^{\rm R} \sigma^0 \tau_1 G^{\rm A}  - \sigma^z \tau_1 G^{\rm R} \sigma^z \tau_1 G^{\rm A} \sigma^0 \tau_1 G^{\rm A} \right] 
\nonumber \\
  &= 8 i (2t)^3 \sum_{\bm k} \sin^2 k_x \cos k_y \cdot {\rm Im} \left[ T^{\rm A} \left\{ (\mu^{\rm R})^2 - (J^{\rm R})^2 + (T^{\rm R})^2 \right\} + 2 T^{\rm R} \left( \mu^{\rm R} \mu^{\rm A} - J^{\rm R} J^{\rm A} \right) \right] . 
\end{align}
Using
\begin{align}
\sum_\k \frac{A_\k}{(D^{\rm R})^2 D^{\rm A}} = -\frac{i\pi \tau^2}{2\mu |\mu|} 
\left[ A_\k \right] - \frac{\pi \tau}{2|\mu|} \left[ A_\k \right]' ,
\label{eq:formula}
\end{align}
where
\begin{align}
\left[ A_\k \right] &= \sum_\k A_\k \delta (\mu^2 - E_\k^2) ,
\\
\left[ A_\k \right]' &= \sum_\k A_\k \frac{\partial}{\partial \mu^2} \delta (\mu^2 - E_\k^2) ,
\end{align}
with $(2\tau)^{-1} = \g = (1 + J^2 / \mu^2) \g_0$, we obtain
\begin{align}
\sigma_{xy}^{z,z} = -\frac{e t^2 \nu}{\mu} \left(\mu^2 - \frac{3}{4}J^2 \right) \tau^2 \left( 1- \frac{J^2}{\mu^2} \right) \left\langle 1 - \cos k_x \cos k_y \right\rangle_{\rm FS} 
\, {\bm n} \cdot ( \partial_x {\bm n} \times \partial_y {\bm n} )  . 
\end{align}
 This contains a term independent to $J$ (which survives the limit $J \to 0$).

 Next, the nonadiabatic terms [Eq.~(\ref{eq:thc-nad})] are calculated as 
\begin{align}
\sigma_{xy}^{z,\perp} 
&= \frac{e}{16 \pi} \left( {\bm A}_{x}^\perp \times {\bm A}_y^\perp \right)^z L 
\label{eq:nonadiabatic}
\\ 
L &= 8 (2t)^3 \sum_{\bm k} \sin^2 k_x \cos k_y \cdot {\rm Im} \left[ T^{\rm A} \left\{ (\mu^{\rm R})^2 + (J^{\rm R})^2 + (T^{\rm R})^2 \right\} + 2  T^{\rm R} \mu^{\rm R} \mu^{\rm A} \right] .
\end{align}
 Similar manipulation leads to 
\begin{align}
 \sigma_{xy}^{z,\perp} = \frac{e t^2 \nu}{\mu} \left(\mu^2 + \frac{1}{4}J^2 \right) \tau^2 
       \left( 1- \frac{J^2}{\mu^2} \right) \left\langle 1 - \cos k_x \cos k_y \right\rangle_{\rm FS} 
 \, {\bm n} \cdot ( \partial_x {\bm n} \times \partial_y {\bm n} )  . 
\end{align} 
Here we also find a $J$-independent term.

 Summing the adiabatic and nonadiabatic terms, the VC-free contribution is obtained as
\begin{align}
\sigma_{\rm SH}^{(0)} &= \sigma_{xy}^{z,z}  + \sigma_{xy}^{z,\perp}  
\nonumber \\
&= \frac{e t^2 \nu}{\mu} (J \tau)^2 \left( 1- \frac{J^2}{\mu^2} \right) \left\langle 1 - \cos k_x \cos k_y \right\rangle_{\rm FS} 
 \, {\bm n} \cdot ( \partial_x {\bm n} \times \partial_y {\bm n} )  .
\label{eq:HC}
\end{align}
 This is Eq.~(7) in the main text. Note that the $J$-independent term is now absent; those in $\sigma_{xy}^{z,z}$ and $\sigma_{xy}^{z,\perp}$ have been canceled. This fact clearly shows the importance of nonadiabatic processes.

\section{Calculation of TSHC with VC}
\label{sec:VCtems}

\begin{figure}[h]
 \includegraphics[width=150mm]{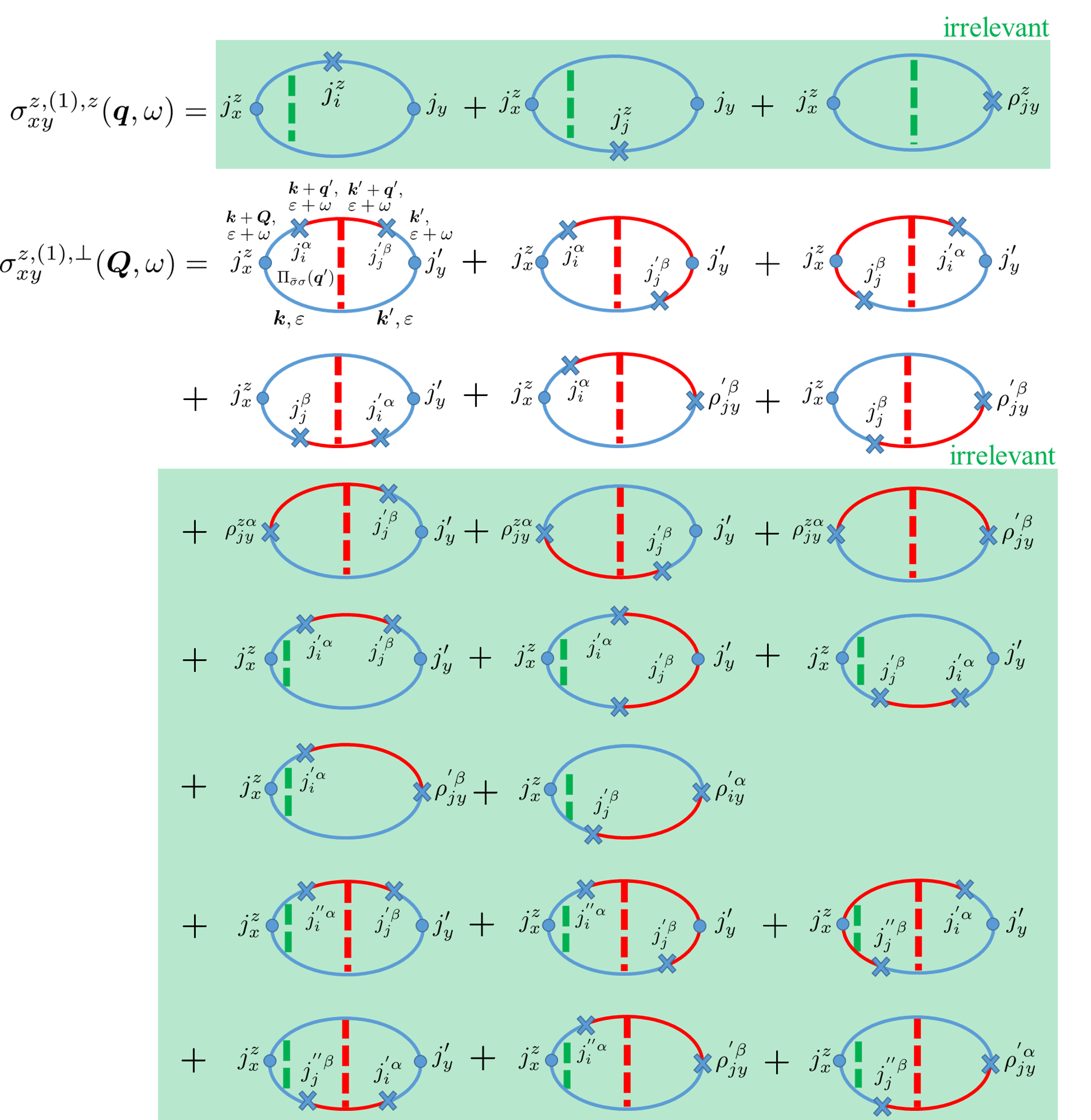}
 \caption{
Feynman diagrams for the TSH conductivity which contain vertex corrections. 
 The red and green broken lines are the longitudinal and transverse spin diffusion propagators, respectively.  
 The primes on $\rho_{jy}^{' \a}$, $j_i^{'' \a}$, etc. are used to distinguish loop momenta. 
The upside-down diagrams are also considered in the calculation. 
 The diagrams in the green-shaded regions are disregarded.}
 \label{fig:wvc}
\end{figure}

 We next consider the effect of vertex corrections (VC). The terms with a longitudinal spin diffusion propagator (green broken line in Fig. \ref{fig:wvc}) with same-spin electrons vanish for the uniform and DC component of spin Hall current. This can be justified by taking the $\Q \to {\bm 0}$ limit before taking the $\omega \to 0$ limit, similar to the case of ferromagnets \cite{NBK}. 

Thus, we consider the nonadiabatic contribution with transverse-spin diffusion propagator, Eq.~(\ref{eq:VC}). Assuming a very slow spatial variation of the texture, we set $q=0$ in Eq.~(\ref{eq:VC}). As in the preceding section, the diagrams that contain $\rho_{ij}^{\a \b}$ vanish, and it suffices to consider 
\begin{align}
\sigma_{xy}^{z,(1)} = -\frac{e}{4\pi} \cdot \frac{1}{4} A_i^\alpha A_j^\beta  \cdot \frac{1}{16}
\left( P + \bar{P} \right)_{ix}^{\alpha,a} \Pi^{ab} \left[ (R + \bar{R}) - (S + \bar{S}) \right]_{jy}^{\beta,b}  , 
\label{eq:wvc}
\end{align}
where $\a, \b = x, y$, and 
\begin{align}
P_{ix}^{\a,a} = \sum_\k {\rm tr} 
\left[ G^{\rm A} j_x^z G^{\rm R} j_i^\a G^{\rm R} \l^a \right] ,
\\
\bar{P}_{ix}^{\a,a} = \sum_\k {\rm tr} 
\left[ G^{\rm A} j_i^\a G^{\rm A} j_x^z G^{\rm R} \l^a \right],
\\
R_{jy}^{\b,b} = \sum_\k {\rm tr} 
\left[ G^{\rm R} j_j^\b G^{\rm R} j_y G^{\rm A} \l^b \right] ,
\\
\bar{R}_{jy}^{\b,b} = \sum_\k {\rm tr} 
\left[ G^{\rm R} j_y G^{\rm A} j_j^\b G^{\rm A} \l^b \right] ,
\\
S_{jy}^{\b,b} = \sum_\k {\rm tr} 
\left[ G^{\rm R} j_y G^{\rm R} j_j^\b G^{\rm A} \l^b \right] ,
\\
\bar{S}_{jy}^{\b,b} = \sum_\k {\rm tr} 
\left[ G^{\rm R} j_i^\b G^{\rm A} j_y G^{\rm A} \l^b \right] .
\end{align}
 Here, $\Pi^{ab}$ are matrix elements of the vertex correction, with indices $a ,b$ taking $ (\g \delta)$, where $\g = x, y$ and $\delta = 0,1,2,3$, which specify the channel $\l^{(\g \delta)} \equiv \sigma^\g \otimes \tau_\delta$. Taking the trace in $\left( P + \bar{P} \right)_{ix}^{\alpha,a}$, $\left( R + \bar{R} \right)_{jy}^{\beta,b}$ and $\left( S + \bar{S} \right)_{jy}^{\beta,b}$, we have 
\begin{align}
\left( P + \bar{P} \right)_{ix}^{\alpha,(\g 0)} 
&= -8 (2t)^2 \delta_{ix} \varepsilon^{\a \g} P_0,
\\
\left( R + \bar{R} \right)_{jy}^{\beta,(\g 0)}
&= 8 (2t)^2 \delta_{jy} \delta^{\g \b} R_0,
\\
\left( S + \bar{S} \right)_{jy}^{\beta,(\g 0)}
&= 8 (2t)^2 \delta_{jy} \delta^{\g \b} S_0,
\\
\left( P + \bar{P} \right)_{ix}^{\alpha,(\g 3)} 
&= 8 (2t)^2 \delta_{ix} \delta^{\a \g} P_3,
\\
\left( R + \bar{R} \right)_{jy}^{\beta,(\g 3)} 
&= 8 (2t)^2 \delta_{jy} \varepsilon^{\g \b} R_3, 
\\
\left( S + \bar{S} \right)_{jy}^{\beta,(\g 3)} 
&= - 8 (2t)^2 \delta_{jy} \varepsilon^{\g \b} S_3, 
\end{align}
with 
\begin{align}
P_0 
&=
\sum_\k \sin^2 k_x  \, 
  {\rm Im} \left[ \mu^{\rm A} \left\{ (\mu^{\rm R})^2 + (J^{\rm R})^2 + (T^{\rm R})^2 \right\} - 2 J^{\rm R} J^{\rm A} \mu^{\rm R} + 2  T^{\rm R} T^{\rm A} \mu^{\rm R} \right] ,
\label{eq:P0}
\\
R_0
&=
\sum_\k \sin^2 k_y \, 
  {\rm Re} \left[ \mu^{\rm A} \left\{ (\mu^{\rm R})^2 + (J^{\rm R})^2 + (T^{\rm R})^2 \right\} - 2 J^{\rm R} J^{\rm A} \mu^{\rm R} + 2  T^{\rm R} T^{\rm A} \mu^{\rm R} \right] ,
\label{eq:R0}
\\
S_0
&=
\sum_\k \sin^2 k_y \, 
  {\rm Re} \left[ \mu^{\rm A} \left\{ (\mu^{\rm R})^2 - (J^{\rm R})^2 + (T^{\rm R})^2 \right\} + 2 T^{\rm R} T^{\rm A} \mu^{\rm R} \right] ,
\label{eq:S0}
\\
P_3 
&= 
\sum_\k \sin^2 k_x \, 
  {\rm Re} \left[  J^{\rm A} \left\{  ( \mu^{\rm R} )^2 + ( J^{\rm R} )^2 + ( T^{\rm R} )^2 \right\} - 2 \mu^{\rm R} \mu^{\rm A} J^{\rm R} \right] ,
\\
R_3 
&=
\sum_\k \sin^2 k_y \, 
 {\rm Im} \left[  J^{\rm A} \left\{  ( \mu^{\rm R} )^2 + ( J^{\rm R} )^2 + ( T^{\rm R} )^2 \right\} - 2 \mu^{\rm R} \mu^{\rm A} J^{\rm R} \right] ,
\\
S_3 
&=
\sum_\k \sin^2 k_y \, 
  {\rm Im} \left[  J^{\rm A} \left\{  ( \mu^{\rm R} )^2 - ( J^{\rm R} )^2 + ( T^{\rm R} )^2 \right\} - 2 T^{\rm R} T^{\rm A} J^{\rm R} \right] .
\end{align}
 Using Eq. (\ref{eq:formula}), we obtain 
\begin{align}
P_0 &= -\pi \nu \tau^2 \frac{\mu^2 - J^2}{\mu^2} \left\langle \sin^2 k_x \right\rangle_{\rm FS} ,
\\
R_0 - S_0 &= - \frac{ \pi \nu \tau }{2\mu} \frac{J^2}{ \mu^2 + J^2 } \left\langle \sin^2 k_y \right\rangle_{\rm FS} .
\end{align}
 Therefore, Eq. (\ref{eq:wvc}) becomes 
\begin{align}
\sigma_{\rm SH}^{(1)} &= 
 \frac{4et^4}{\pi}   ({\bm A}_x \times {\bm A}_y)^z  \cdot 
 P_0 \, \Pi_{\bar\sigma \sigma}  (R_0 - S_0)  
\label{eq:decomp}
\\
&= \frac{2 e t^2 \nu}{\mu} 
(J\tau)^2
\frac{(2 t)^2}{\mu^2 + J^2} 
\frac{\tau^{-1}}{ \tau_\varphi^{-1} + \tau_{\rm s}^{-1} }  
\left\langle
\sin^2 k_x
 \right\rangle_{\rm FS}^2   \, {\bm n} \cdot ( \partial_x {\bm n} \times \partial_y {\bm n} ) . 
\end{align}
 This is Eq. (8) in the main text. We note that the contribution with $\Pi^{(\g 3)(\g 3)}$ is smaller than the one with $\Pi^{(\g 0)(\g 0)}$ by a factor of $(J/\mu^2 \tau)^2$, and is zeroth order in $\tau$; thus it can be disregarded for good metals considered here.

\begin{figure}[b]
 \includegraphics[width=170mm]{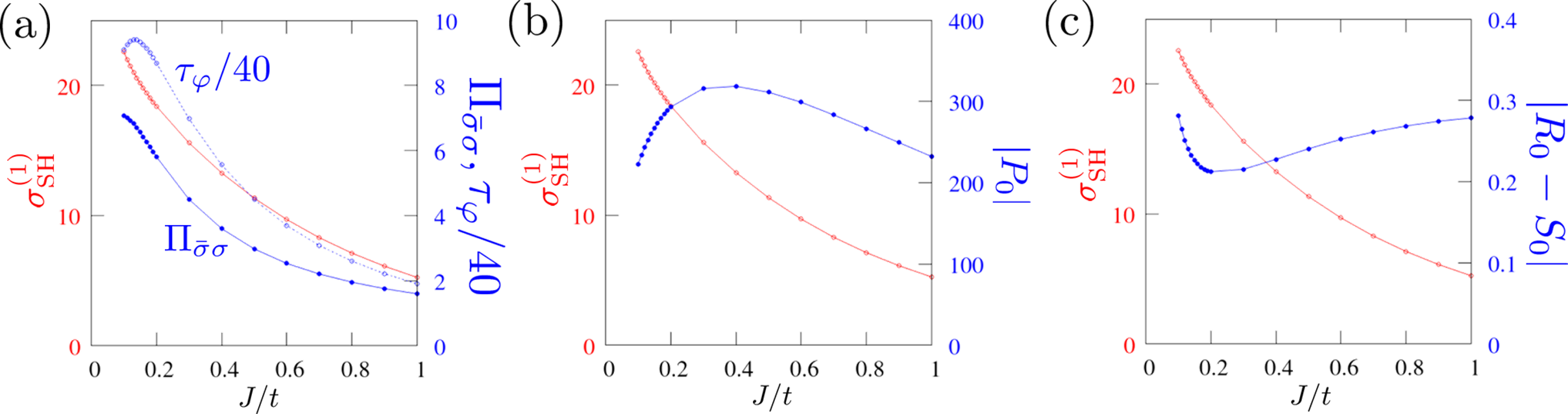}
 \caption{
 Comparison of $J$-dependence of the peak value of $\sigma_{\rm SH}^{(1)}$ (red lines) 
with (a) $\Pi_{\bar\sigma \sigma}$ and $\tau_\varphi$, (b) $|P_0|$, and (c) $| R_0 - S_0|$ 
(blue lines). 
 See Eqs.~(\ref{eq:VC}) and (\ref{eq:P0})-(\ref{eq:S0}) for the definition 
of $\Pi_{\bar\sigma \sigma}$, $P_0$, $R_0$, and $S_0$, 
and Eq.~(\ref{eq:decomp}) for the relation among them. 
 In the plots, we set $e$, $t$, and the lattice constant unity, 
and $\tau_\varphi$ is divided by 40. 
}
 \label{fig:J-dep}
\end{figure}

\section{Enhancement at weak coupling}
\label{sec:J-dep}

 As discussed in the main text, one of the most spectacular aspect of the present TSHE is the enhancement at weak coupling (small $J$). More specifically, the peak value of the spin Hall conductivity near the AF gap edge increases as $J$ is decreased. To see the origin of this enhancement, we compare in Fig.~\ref{fig:J-dep} the $J$-dependence of the peak value of $\sigma_{\rm SH}^{(1)}$ with that of $\Pi_{\bar{\s}\s}$ (vertex correction; see Eq.~(\ref{eq:VC})), $\tau_\varphi$ (spin dephasing time), $P_0$ (left loop integral), and $R_0 - S_0$ (right loop integral) [see Eq.~(\ref{eq:decomp})] in the weak coupling regime, $J/t<1$. The value of $\mu$ is taken at which $\sigma_{\rm SH}^{(1)}$ is peaked. As seen, the overall $J$-dependence of $\sigma_{\rm SH}^{(1)}$ follows that of $\Pi_{\bar\sigma \sigma}$ and $\tau_\varphi$, rather than $P_0$ and $R_0 - S_0$, implying that the spin dephasing time is the key parameter for the enhancement of TSHE. At very small $J/t$ ($ < 0.2$), $|R_0 - S_0|$ shows a remarkable increase, which may also be important. Overall, it is very important to consider $\tau_\varphi$ in the weak coupling regime.

\section{Perturbative treatment of $J$}
\label{sec:pert}

\begin{figure}[b]
 \includegraphics[width=150mm]{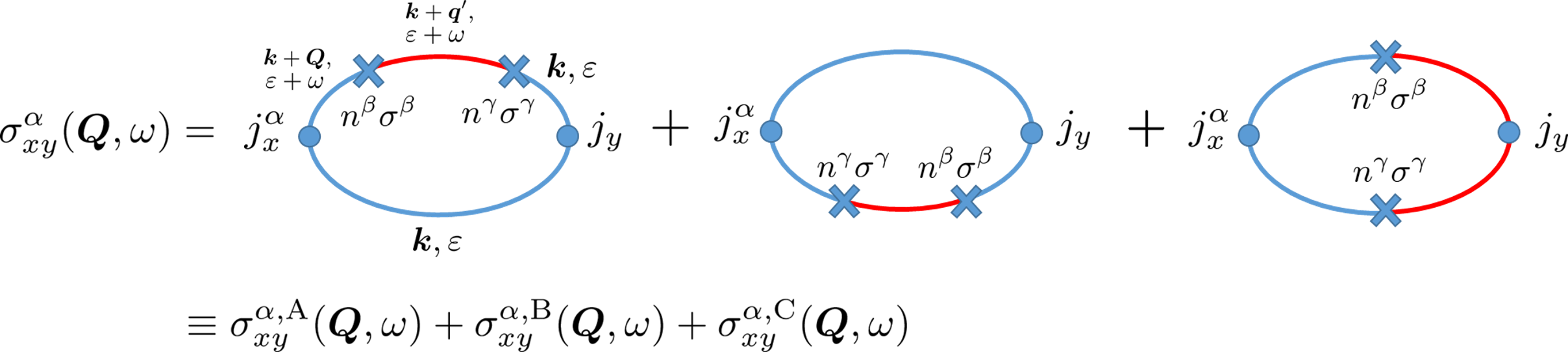}
 \caption{ Feynman diagrams for the TSH conductivity at second order in $J$.}
 \label{fig:pert}
\end{figure}

 In this section, we calculate the TSH conductivity by treating the exchange coupling $J$ perturbatively. 
 Relevant diagrams are shown in Fig.~\ref{fig:pert}. 
 The Green's functions in this section are those at $J=0$. 
 Extracting $\q$ to the lowest (second) order, we write the antisymmetric part as 
\begin{align}
\sigma_{xy}^{\a,{\rm A}} &= -\frac{e}{8\pi} J^2 q_x q'_y n_\q^\b n_{\q'}^\g (-2t)^3
\sum_\k \cos k_x \sin^2 k_y {\rm tr} \left[ \s^\a \s^\b \s^\g \otimes \tau_1 G^{\rm R} \tau_3 G^{\rm R} \tau_1 G^{\rm R} \ \tau_3 G^{\rm R} \tau_1 G^{\rm A} \right],
\\
\sigma_{xy}^{\a,{\rm B}} &= -\frac{e}{8\pi} J^2 q'_x q_y n_\q^\b n_{\q'}^\g (-2t)^3 
\sum_\k \cos k_x \sin^2 k_y {\rm tr} \left[ \s^\a \s^\b \s^\g \otimes \tau_1 G^{\rm R} \tau_1 G^{\rm A} \tau_3 G^{\rm A} \tau_1 G^{\rm A} \tau_3 G^{\rm A} \right], 
\\
\sigma_{xy}^{\a,{\rm C1}} &= \frac{e}{8\pi} J^2 q'_x q_y n_\q^\b n_{\q'}^\g (-2t)^3 
\sum_\k \cos k_x \sin^2 k_y {\rm tr} \left[ \s^\a \s^\b \s^\g \otimes \tau_1 G^{\rm R} \tau_1 G^{\rm R} \tau_3 G^{\rm R} \tau_1 G^{\rm A} \tau_3 G^{\rm A} \right], 
\\
\sigma_{xy}^{\a,{\rm C2}} &= \frac{e}{8\pi} J^2 q_x q'_y n_\q^\b n_{\q'}^\g (-2t)^3 
\sum_\k \cos k_x \sin^2 k_y {\rm tr} \left[ \s^\a \s^\b \s^\g \otimes \tau_1 G^{\rm R} \tau_3 G^{\rm R} \tau_1 G^{\rm A} \tau_3 G^{\rm A} \tau_1 G^{\rm A}  \right] , 
\end{align}
where $\sigma_{xy}^{\a, {\rm C1}} + \sigma_{xy}^{\a, {\rm C2}} = \sigma_{xy}^{\a, {\rm C}}$. 
 Noting $\left( \mu^{\rm R(A)} \right)^2 - \left( T^{\rm R(A)} \right)^2 = 1/D^{\rm R(A)} $ 
and Eq. (\ref{eq:formula}), we proceed as 
\begin{align}
\sigma_{xy}^{\a,{\rm A}} + \sigma_{xy}^{\a, {\rm B}} 
&= \frac{i e}{2\pi} J^2 \left( \partial_x {\bm n} \times \partial_y {\bm n} \right)^\a (-2t)^3
2i {\rm Im} \sum_\k \cos k_x \sin^2 k_y \left[ (\mu^{\rm R})^2 - (T^{\rm R})^2 \right]^2 T^{\rm A} 
\\
&= -\frac{e}{\pi} J^2 \left( \partial_x {\bm n} \times \partial_y {\bm n} \right)^\a (-2t)^3  
{\rm Im} \sum_\k \frac{\cos k_x \sin^2 k_y T_\k }{ \left( D^{\rm R} \right)^2 D^{\rm A} } 
\\
&= \frac{1}{2} \frac{e t^2 \nu}{\mu} (J\tau)^2 \left\langle 1 - \cos k_x \cos k_y \right\rangle_{\rm FS} \left( \partial_x {\bm n} \times \partial_y {\bm n} \right)^\a , 
\\
\sigma_{xy}^{\a,{\rm C1}} + \sigma_{xy}^{\a,{\rm C2}}
&= \frac{ie}{2\pi} J^2 \left( \partial_x {\bm n} \times \partial_y {\bm n} \right)^\a (-2t)^3
2i {\rm Im} \sum_\k \cos k_x \sin^2 k_y \left[ (\mu^{\rm R})^2 - (T^{\rm R})^2 \right] \left[ (\mu^{\rm A})^2 - (T^{\rm A})^2 \right] T^{\rm R} 
\\
&= -\frac{e}{\pi} J^2 \left( \partial_x {\bm n} \times \partial_y {\bm n} \right)^\a (-2t)^3 
{\rm Im} \sum_\k \frac{\cos k_x \sin^2 k_y T_\k }{ \left( D^{\rm R} \right)^2 D^{\rm A} } 
\\
&= \frac{1}{2} \frac{e t^2 \nu}{\mu} (J\tau)^2 \left\langle 1 - \cos k_x \cos k_y \right\rangle_{\rm FS} \left( \partial_x {\bm n} \times \partial_y {\bm n} \right)^\a . 
\end{align} 
 Therefore, we obtain 
\begin{align}
j_{{\rm s},x}^\a = \frac{e t^2 \nu}{\mu} (J\tau)^2  \left\langle 1 - \cos k_x \cos k_y \right\rangle_{\rm FS} \left( \partial_x {\bm n} \times \partial_y {\bm n} \right)^\a E_y . 
\end{align}
 This is Eq.~(10) in the main text, and agrees with Eq.~(\ref{eq:HC}) at order $J^2$.

\section{Spin Hall angle}
\label{sec:VCtems}

\begin{figure}[h]
 \includegraphics[width=150mm]{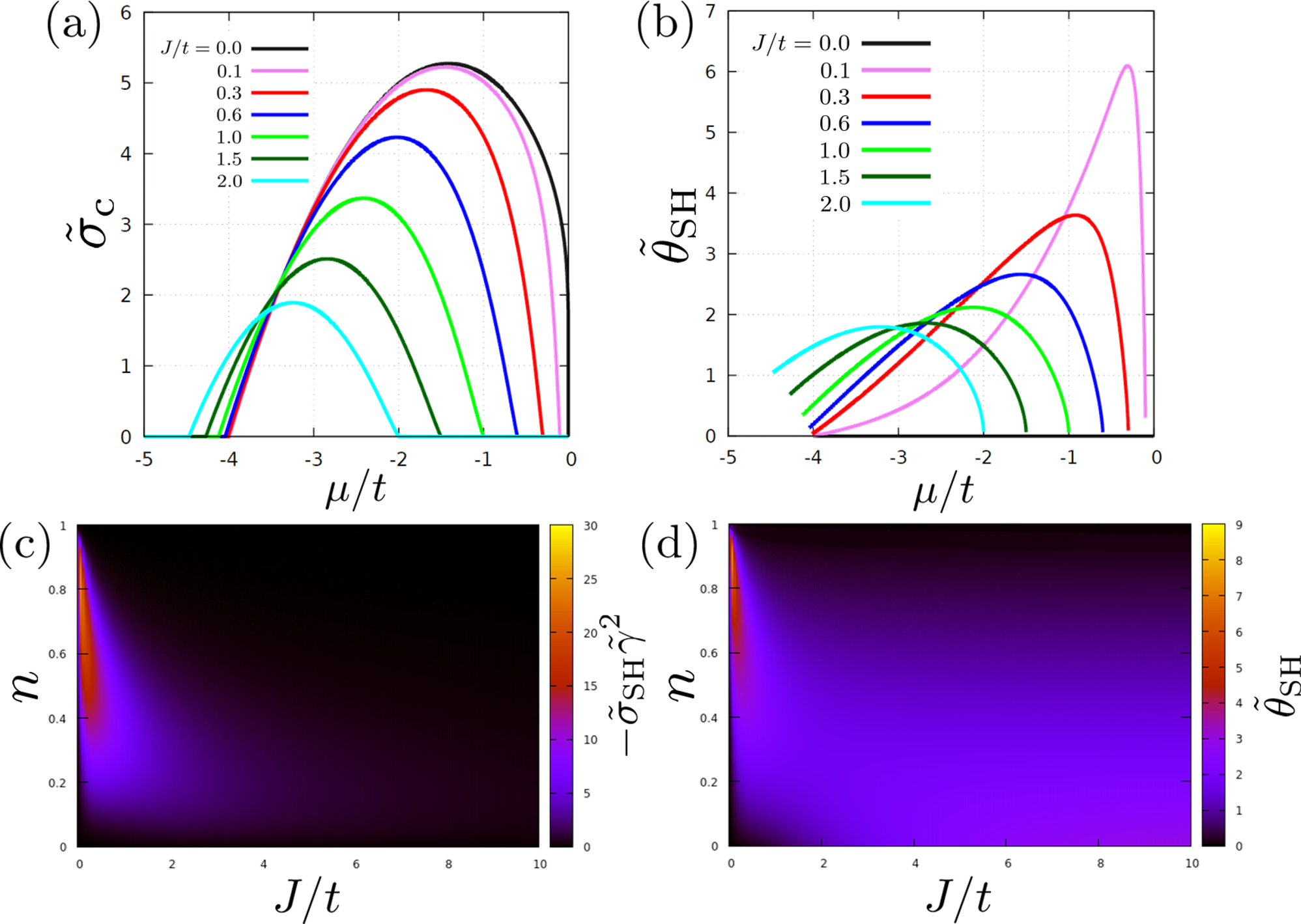}
 \caption{
(a) and (b) show the chemical potential dependence of the normalized longitudinal conductivity (a) and the spin Hall angle (b) for several choices of $J/t$. 
(c) and (d) are color plots in the plane of $n$ (average electron number per site) and $J/t$ of the normalized topological spin Hall conductivity (c) and the normalized spin Hall angle (d). ($n=1$ corresponds to completely filled lower and empty upper AF bands.) 
}
 \label{fig:cond_angle}
\end{figure}

 For a comparison to experiments, we plot a normalized spin Hall angle $\tilde{\theta}_{\rm SH} \equiv \tilde{\sigma}_{\rm SH} / \tilde{\sigma}_{\rm c}$, where $\tilde{\sigma}_{\rm c} = \sigma_{\rm c} \tilde{\gamma}$ ($\sigma_{\rm c} = 2 e^2 D\nu$) is the normalized longitudinal conductivity [Fig. \ref{fig:cond_angle}(a)]. 
 
 The spin Hall angle $\tilde{\theta}_{\rm SH}$ is shown in Fig.~\ref{fig:cond_angle}(b). As $J$ is increased from $J=0.1t$, the peak value decreases first quickly, then moderately and becomes stationary. This is because the longitudinal conductivity also decreases but rather constantly; see Fig.~\ref{fig:cond_angle}(a).

 The normalized TSH conductivity $\tilde{\sigma}_{\rm SH} \tilde \gamma^2$ and the normalized TSH angle $\tilde{\theta}_{\rm SH}$ are plotted in Fig.~\ref{fig:cond_angle} (c) and (d), respectively, in the plane of $J$ and electron filling $n$. As for the former [Fig.~\ref{fig:cond_angle}(c)], we see a very sharp peak in a very narrow region at weak coupling ($J \lesssim 0.5t$) and $n \gtrsim 0.4$. This is discussed in the main text. The spin Hall angle $\tilde{\theta}_{\rm SH}$ [Fig.~\ref{fig:cond_angle}(d)] has a similar structure, but also has a broad tail (or plateau) extended to larger $J$.

\section{nonlocality}

 As the wave-vector ${\bm q}$ of spin texture (or spin gauge field) is increased, the effective field becomes nonlocal with respect to the spin texture. This is induced through spin diffusion. To study this, we retain the ${\bm q}$-dependence of the texture, and write the TSH conductivity as
\begin{align}
\sigma_{\rm SH} (\Q) \propto \Pi (\q') [ {\bm A}_x^\perp (\q) \times {\bm A}_y^\perp (\q') ]^z 
  + (\q \leftrightarrow \q') - (x \leftrightarrow y). 
\end{align}
 In real space, it reads 
\begin{align}
 \sigma_{\rm SH} 
&\propto  \int d\r' \Pi( \r - \r' ) [{\bm A}_x^\perp (\r) \times {\bm A}_y^\perp (\r')]^z 
  - (x \leftrightarrow y)
\nonumber \\
&= {\bm n} \cdot ( \partial_x {\bm n} \times {\cal R} ( \hat{z} \times \tilde{\bm A}_y^\perp) ) - (x \leftrightarrow y)
\nonumber \\
&= {\bm n} \cdot ( \partial_x {\bm n} \times \tilde{\bm d}_y ) - (x \leftrightarrow y)  , 
\end{align}
with
\begin{align} 
\tilde{\bm A}_i^\perp &= \int d\r' \Pi( \r - \r' ) {\bm A}_i^\perp (\r'), 
\\
\tilde{\bm d}_i &= \int d\r' \Pi (\r - \r') {\cal R} (\r) {\cal R}^{-1} (\r') \partial_i {\bm n} (\r') , 
\end{align}
and   
\begin{align}
\Pi ({\bm R}) 
= \sum_\q \Pi_{\bar\sigma \sigma} (\q) \, \expo^{i \q \cdot {\bm R}}
= \frac{2}{\pi^2 \nu \tau} \frac{\mu^2}{\mu^2 - J^2} \frac{1}{D\tau} K_0 (R/\ell_{\vp {\rm s}}),
\end{align} 
where $\ell_{ \vp {\rm s} } = \sqrt{ D \tau_{\vp {\rm s} }}$ with $\tau_{\vp {\rm s}}^{-1} = \tau_{\vp}^{-1} + \tau_{\rm s}^{-1} $, ${\bm R} =  \r - \r' $, and $K_0 (x)$ is the modified Bessel function of the second kind. We identify the physical spin Hall current as 
\begin{align}
 {\bm j}_{{\rm s},i}
 \propto ( \partial_i {\bm n} \times \tilde{\bm d}_j - \partial_j {\bm n} \times \tilde{\bm d}_i ) E_j . 
\end{align}

\section{Numerics with Landauer-B\"uttiker formula}

\begin{figure}[t]
 \includegraphics[width=140mm]{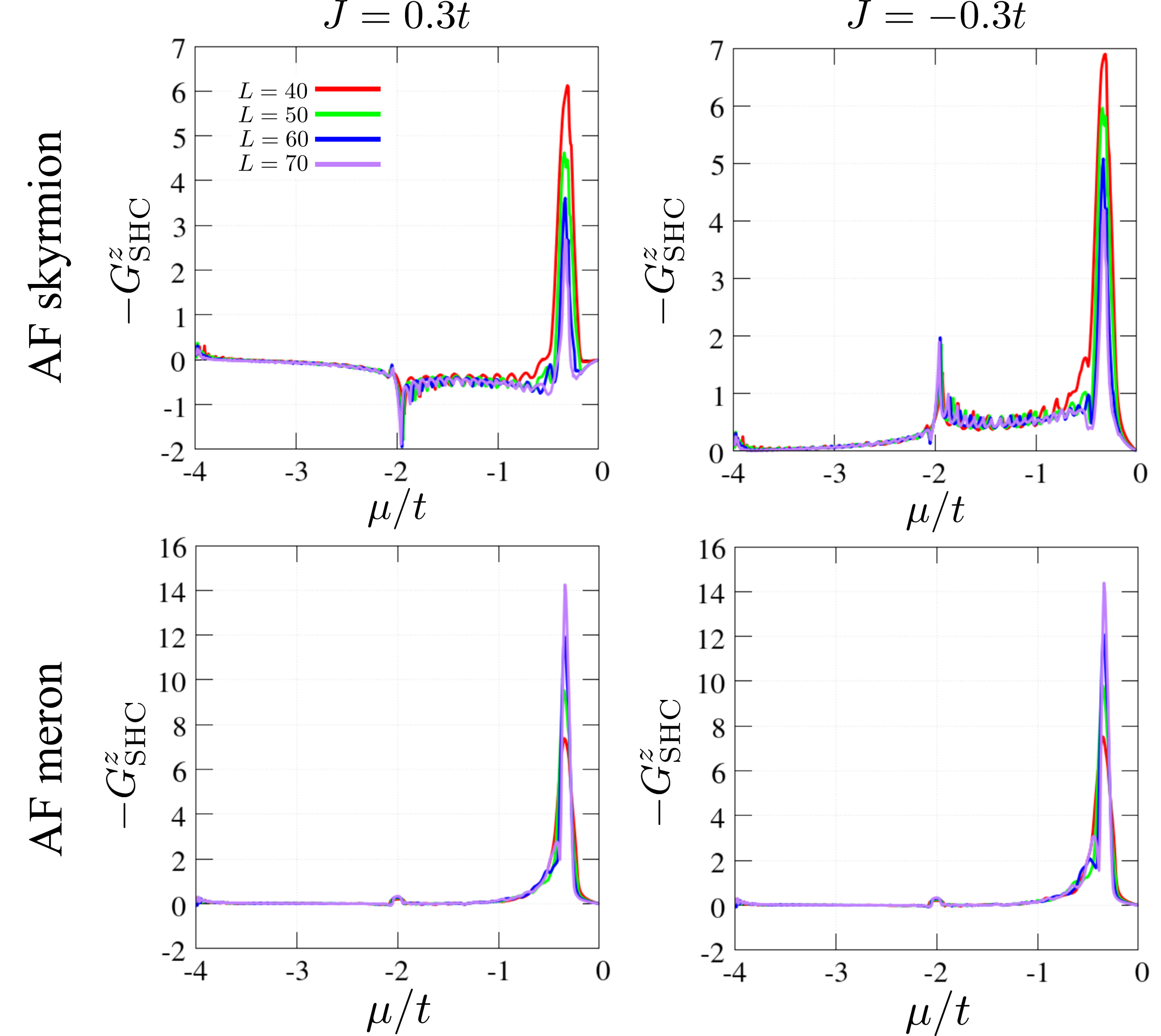}
 \caption{
 Chemical potential dependence of the spin Hall conductance before the \lq\lq symmetrization''  
for systems with AF skyrmion (upper panels) and AF meron (lower panels), 
and for $J = 0.3t$ (left panels) and $J=-0.3t$ (right panels). 
}
 \label{fig:Jodd}
\end{figure}

 In the calculation based on the Landauer-B\"uttiker formula, we employ the four-terminal geometry with nonmagnetic leads and obtained numerically the spin Hall conductance from the transmission coefficients \cite{Ohe}. The magnetic textures employed are as follows. For AF skyrmion, we take 
\begin{align}
{\bm n} (\r) = \left( \cos\varphi \sin\theta, \ \sin\varphi \sin\theta, \ \cos\theta \right) , 
\end{align}
where $\varphi = {\rm Arg}(x + iy) + \pi/2$ and $\theta = \theta_+ + \theta_-$ with $\sin \theta_\pm = \tanh \frac{2(|\r| \pm R_{\rm sk})}{R_{\rm sk}}$ \cite{Romming}. For AF meron, we take
\begin{align}
{\bm n} (\r) &= \left( \cos\varphi \sin\theta, \ \sin\varphi \sin\theta, \ \cos\theta \right)  \quad (\cos \theta > 0)
\\
{\bm n} (\r) &= \left( \cos\varphi, \ \sin\varphi, \ 0 \right) \quad ({\rm otherwise}) , 
\end{align}
with same $\varphi$ and $\theta$.

The results are shown in Fig.~\ref{fig:Jodd} for $J = 0.3t$ and $J = -0.3t$. While the analytical result indicates that contributions from vector chirality are even in $J$ (see the main text), a substantial $J$-odd component is seen in Fig.~\ref{fig:Jodd} for AF skyrmion whereas it is almost absent for AF meron. We have thus extracted the $J$-even contribution by symmetrizing as 
\begin{align}
  G_{\rm SHC}^z \equiv \left[ G_{\rm SHC}^z (J) + G_{\rm SHC}^z (-J) \right] / 2 . 
\end{align} 
 Figure 3 in the main text shows this \lq\lq symmetrized'' data. Since the (disregarded) $J$-odd components do not depend much on the system size, they are likely to come from the boundary between the sample and the lead.


\begin{thebibliography}{9}
\bibitem{SHE0} J. Sinova, S. O. Valenzuela, J. Wunderlich, C.~H. Back, and T. Jungwirth, Rev. Mod. Phys. \textbf{87}, 1213 (2015).
\bibitem{SHE1} M. I. Dyakonov and V. I. Perel, JETP, Lett. \textbf{13}, 467 (1971); Phys. Lett. \textbf{A35}, 459 (1971). 
\bibitem{SHE2} S. Murakami, N. Nagaosa, and S. -C. Zhang, Science \textbf{301}, 1348 (2003). 
\bibitem{SHE3} J. Sinova, D. Culcer, Q. Niu, N. A. Sinitsyn, T. Jungwirth, and A. H. MacDonald, Phys. Rev. Lett. \textbf{92}, 126603 (2004). 
\bibitem{SHE4} Y. K. Kato, R. C. Myers, A. C. Gossard, and D. D. Awschalom, Science \textbf{306}, 1910 (2004).  
\bibitem{SHE5} J. Wunderlich, B. Kastner, J. Sinova, and T. Jungwirth, Phys. Rev. Lett. \textbf{94}, 047204 (2004). 

\bibitem{com1} This applies to the strong $s$-$d$ coupling regime. In the weak-coupling regime, there is a proposal recently that the essence is the Hall effect in the charge channel \cite{Denisov}. 

\bibitem{Denisov} K. S. Denisov, I. V. Rozhansky, N. S. Averkiev, and E. L\"ahderanta, Phys. Rev. Lett. \textbf{117}, 027202 (2016); Phys. Rev. B \textbf{98}, 195439 (2018); I. V. Rozhansky, K. S. Denisov, M. B. Lifshits, N. S. Averkiev, and E. Lähderanta, Phys. Status Solidi (b) 1900033 (2019).   



\bibitem{Menders2014} J. B. S. Mendes, R. O. Cunha, O. Alves Santos, P. R. T. Ribeiro, F. L. A. Machado, R. L. Rodríguez-Suárez, A. Azevedo, and S. M. Rezende \textit{et al.}, Phys. Rev. B \textbf{89}, 140406(R) (2014).


\bibitem{Buhl} P. M. Buhl, F. Freimuth, S. Bl\"ugel, and Y. Mokrousov, Phys. Status Solidi RRL \textbf{11}, 1700007 (2017). 
\bibitem{Gobel} B. G\"obel, A. Mook, J. Henk, I. Mertig, Phys. Rev. B \textbf{96}, 060406(R) (2017). 
\bibitem{Akosa} C.~A. Akosa, O.~A. Tretiakov, G. Tatara, and A. Manchon, Phys. Rev. Lett. \textbf{121}, 097204, (2018). 

\bibitem{com:VNC} One may also think of \lq\lq N\'eel vector chirality'', which, however, has more flavor of scalar chirality formed by the N\'eel vector, i.e., $\n \cdot ( \d_i \n \times \d_j \n )$. 


\bibitem{TT} K. Taguchi and G. Tatara, Phys. Rev. B \textbf{79}, 054423 (2009)
\bibitem{IN} H. Ishizuka and N. Nagaosa, New J. Phys. \textbf{20}, 123027 (2018). 
\bibitem{TG} G. Tatara and N. Garcia, Phys. Rev. Lett. \textbf{91}, 076806 (2003). 
\bibitem{KNB} H. Katsura, N. Nagaosa, and A. V. Balatsky, Phys. Rev. Lett. \textbf{95}, 057205 (2005). 
\bibitem{KKAT} T. Kikuchi, T. Koretsune, R. Arita, and G. Tatara, Phys. Rev. Lett. \textbf{116}, 247201 (2016). 
\bibitem{AFexp3} H. Jani, J. -C. Lin, J. Chen, J. Harrison, F. Maccherozzi, J. Schad, S. Prakash, C. -B. Eom, A. Ariando, T. Venkatesan and P. G. Radaelli, Nature \textbf{590}, 74-79 (2021).


\bibitem{AFSk1} C. Jin, C. Song, J. Wang, and Q. Liu, Appl. Phys. Lett. \textbf{109}, 182404 (2016).
\bibitem{AFSk4} X. Zhang, Y. Zhou, and M. Ezawa, Sci. Rep. \textbf{6}, 24795 (2016). 
\bibitem{AFSk2} H. Velkov, O. Gomonay, M. Beens, G. Schwiete, A. Brataas, J. Sinova, and R.~A. Duine, New J. Phys. \textbf{18}, 075016 (2016). 
\bibitem{AFSk3} J. Barker and O.~A. Tretiakov, Phys. Rev. Lett. \textbf{116}, 147203 (2016). 
\bibitem{AFexp1}  T. Dohi, S. DuttaGupta, S. Fukami, and H. Ohno, Nat. Commun. \text{10}, 5153 (2019). 
\bibitem{AFexp2} W. Legrand, D. Maccariello, F. Ajejas, S. Collin, A. Vecchiola, K. Bouzehouane, N. Reyren, V. Cros, and A. Fert, Nat. Mater. {\bf 19}, 34 (2020). 
\bibitem{AFSk5} R. Yambe and S. Hayami, Phys. Rev B {\bf 107}, 014417 (2023). 


\bibitem{Tatara} G. Tatara, H. Kohno, and J. Shibata, Phys. Rep. \textbf{468}, 213-301 (2008). 
\bibitem{Nakane1} J. J. Nakane, K. Nakazawa, and H. Kohno, Phys. Rev. B {\bf 101}, 174432 (2020).  

\bibitem{NBK} K. Nakazawa, M. Bibes, and H. Kohno, J. Phys. Soc. Jpn. \textbf{87}, 033705 (2018); K. Nakazawa, and H. Kohno, Phys. Rev. B \textbf{99}, 174425, (2019). 

\bibitem{Streda} P. St\v{r}eda, J. Phys. C \textbf{15}, L717 (1982). 

\bibitem{suppl} See Supplemental Material for calculational details, explicit formula of the adiabatic and nonadiabatic contributions, analysis on the enhancement at weak couling, perturbative treatment of $J$, and further plots including the spin Hall angle. 


\bibitem{com:FS}  Explicitly, they are evaluated as \cite{Nakane1} 
\begin{align}
  C_{xy} &= \frac{1}{4\pi^2} \frac{|\mu|}{t^2 \nu}
\left[ \frac{2}{\tilde \mu} K(x) - \left( 1 + \frac{1}{\tilde \mu} \right) E(x)  \right],
\nonumber \\
  C_{xx} &= \frac{1}{4\pi^2} \frac{|\mu|}{t^2 \nu} \left[ \left(1 + \frac{1}{\tilde \mu} \right) E(x) - 2 K(x) \right], 
\nonumber 
\end{align}
where $K(x)$ and $E(x)$ are the complete elliptic integrals of the first and the second kind, respectively, and $\tilde \mu = \sqrt{\mu^2 - J^2}/4t$ and $x = (1 - \tilde{\mu})/(1 + \tilde{\mu})$ are dimensionless parameters. 


\bibitem{Manchon} A. Manchon, J. Phys. Condens. Matter \textbf{29}, 104002 (2017). Their $\beta$ seems to be $\sin \chi_{k_{\rm F}}$ instead of $\cos \chi_{k_{\rm F}}$.  

\bibitem{com0} A microscopic calculation of the spin relaxation rate in AF metal can be found in Ref.~\cite{Nakane2}, in which magnetic impurities are considered as a spin sink. 

\bibitem{com2} The nonadiabatic process is important at a more fundamental level. Both the adiabatic and nonadiabatic contributions to $\tilde \sigma_{\rm SH}^{(0)}$ contain terms of $O(J^0)$, which do not vanish at $J=0$, but when added they cancel each other and the resulting $\tilde \sigma_{\rm SH}^{(0)}$ vanishes at $J=0$ \cite{suppl}. 

\bibitem{HLN} S. Hikami, A. I. Larkin, and Y. Nagaoka, Prog. Theor. Phys. \textbf{63}, 707 (1980).

\bibitem{Nakane2} J. J. Nakane and H. Kohno, Phys. Rev. B {\bf 103}, L180405 (2021).  

\bibitem{Tatara2002} G. Tatara and H. Kawamura, J. Phys. Soc. Jpn. {\bf 71}, 2613 (2002). 

\bibitem{Ohe2017} J. Ohe, T. Ohtsuki, and B. Kramer, Phys. Rev. B {\bf 75},  245313 (2017).


\bibitem{Kohno2007} H. Kohno and J. Shibata, J. Phys. Soc. Jpn. {\bf 76}, 063710 (2007). 

\bibitem{Meshcheriakova} O. Meshcheriakova, S. Chadov, A. K. Nayak, U. K. R\"o\ss ler, J. K\"ubler, G. Andr\'e, A. A. Tsirilin, J. Kiss, S. Hausdorf, A. Kalache, W. Schnelle, M. Nicklas, and C. Felser, Phys. Rev. Lett. {\bf 113}, 087203 (2014). 

\bibitem{Vistoli} L. Vistoli, W. Wang, A. Sander, Q. Zhu, B. Casals, R. Cichelero, A. Barth\'el\'emy, S. Fusil, G. Herranz, S. Valencia, R. Abrudan, E. Weschke, K. Nakazawa, H. Kohno, J. Santamaria, W. Wu, V. Garcia, and M. Bibes, Nat. Phys. \textbf{15}, 67 (2019).


\end{thebibliography}

\begin{thebibliography}{9}
\bibitem{Nakane} J. J. Nakane, K. Nakazawa, and H. Kohno, Phys. Rev. B {\bf 101}, 174432 (2020). 
\bibitem{Manchon} A. Manchon, J. Phys. Condens. Matter \textbf{29}, 104002 (2017). 
\bibitem{NBK} K. Nakazawa, M. Bibes, and H. Kohno, J. Phys. Soc. Jpn. \textbf{87}, 033705 (2018); K. Nakazawa and H. Kohno, Phys. Rev. B \textbf{99}, 033705 (2018). 
\bibitem{Ohe} J. Ohe, T. Ohtsuki, B. Kramer, Phys. Rev. B {\bf 75}, 245313 (2007). 
\bibitem{Romming} N. Romming, A. Kubetzka, C. Hanneken, K. von Bergmann, and R. Wiesendanger, 
Phys. Rev. Lett. {\bf 114}, 177203 (2015). 
\end{thebibliography}
\end{document}